\documentclass[showkeys,superscriptaddress,twocolumn,pre,showpacs,preprintnumbers]{revtex4-1}
\usepackage{amsmath,amssymb}
\usepackage{epsfig,graphics,color,calc,graphicx}
\usepackage{tikz}
\usepackage{mathrsfs}

\newcommand{\newatop}[2]{\genfrac{}{}{0pt}{}{#1}{#2}}

\newcommand{\rr}[1]{{\normalfont\textrm{#1}}}

\newcommand{\parametro}{\textbf}
\newcommand{\fcar}{{\mathbb{I}}}

\newlength{\pecettawidth}
\begin{document}
\title{Folding transitions in three--dimensional space with defects}

\author{Emilio N.M.\ Cirillo}
\email{emilio.cirillo@uniroma1.it}
\thanks{corresponding author.}
\affiliation{Dipartimento di Scienze di Base e Applicate per
             l'Ingegneria, Sapienza Universit\`a di Roma,
             via A.\ Scarpa 16, I--00161, Roma, Italy.}

\author{Giuseppe Gonnella}
\email{Giuseppe.Gonnella@ba.infn.it} \affiliation{Dipartimento di
Fisica dell'Universit\`a degli Studi di Bari, {\rm and} INFN,
Sezione di Bari,  via Amendola 173, 70126 Bari, Italy.}

\author{Alessandro Pelizzola}
\email{alessandro.pelizzola@polito.it}
\affiliation{Dipartimento di Scienza Applicata e Tecnologia, CNISM and Center for Computational
Studies, Politecnico di Torino,
Corso Duca degli Abruzzi 24, I--10129 Torino, Italy}
\affiliation{INFN, Sezione di Torino, via Pietro Giuria 1, I-10125 Torino,
Italy}
\affiliation{Human Genetics Foundation, HuGeF, Via Nizza 52, I-10126 Torino,
Italy}


\begin{abstract}
  A model describing the three--dimensional folding of the triangular
  lattice on the face--centered cubic lattice is generalized allowing
  the presence of defects, which are related to cuts in the
  two--dimensional network. The model can be expressed in terms of
  Ising--like variables with nearest--neighbor and plaquette
  interactions in the hexagonal lattice; its phase diagram is
  determined by means of the
Cluster Variation Method.
The results found by varying the curvature and defect energy show
that the introduction of defects turns the first--order crumpling
transitions of the model without defects into continuous
transitions. New phases also appear by decreasing the  energy cost
of defects and the behavior of their  densities  has been analyzed.

\end{abstract}





\maketitle

\section{Introduction}
\label{s:introduzione}
\par\noindent

The behavior of fluctuating membranes and surfaces is relevant for
many physical and biological systems, from gauge theories and
strings to  vesicles and cellular membranes; see Refs.\
\cite{NPW,GDZ} for reviews.  A  class is given by   polymerized or
crystalline membranes \cite{KKN,BowickPhysRep,DFGPhysRep} consisting
of two--dimensional networks of molecules with fixed connectivity.
Examples are the spectrin network in red--blood cells \cite{ACT} or
the graphene \cite{GRA}. Theoretical arguments suggest the relevance
of curvature energy terms for the macroscopic behavior of
fluctuating surfaces \cite{NPW}. For polymerized membranes, at
variance with fluid membranes \cite{PL}, a flat phase with
long--range order in the orientation of the normals to the surface
is expected to be stable at high bending rigidity \cite{NP}.

The prediction of a crumpling transition for {\it phantom}
polymerized membranes (where self--avoidance is not taken into
account) is confirmed on the basis of various analytical results
\cite{DG,PKN,AL,LR} and  numerical simulations \cite{KKN,KN,BEW} on
continuous models. The character of the transition is controversial.
Among the most recent results, the simulations of Refs.\ \cite{KD}
suggest a first--order behavior while the non--perturbative
renormalization group calculations of \cite{KM} predict  for a
phantom membrane embedded in the three--dimensional space  a
continuous  transition.

A different approach has been to study discrete models for
polymerized membranes whose nodes are constrained to occupy
positions corresponding to the sites of a given lattice. Solutions
of discrete models, as in other domains of statistical physics, can
represent a reference for the behavior of fluctuating membranes of a
given class. It is well--known that Ising--like models
\cite{ccgm,cgjp} naturally admit an interpretation in terms of a
surface gas with a variable number of components. More difficult is
the problem of representing a single membrane in terms of discrete
variables. In Ref.\ \cite{npb95} Bowick and co--workers have been
able to  model  the folding of a triangular network in the
face--centered cubic (fcc) lattice in terms of local discrete
variables, subject to local constraints. Since the distance between
the nodes of the network is fixed, in this model one can consider as
frozen  the ``phonon" degrees of freedom of the membrane while  only
the bending modes are taken into account \cite{KJ}. A simpler
version of the folding problem of \cite{npb95} has been studied in
\cite{epl,pre,nostro,Nishi4} corresponding to the case of a
two--dimensional embedding space. Here the normals to the triangles
of the network can point only ``up'' or ``down'' i n some direction.
A similar model concerning the folding properties of a square
lattice along the main axis and the diagonals has been studied in
\cite{DiF,NostroNPB}. Finally, the three--dimensional  folding
problems of the triangular lattice with quenched random bending
rigidity and spontaneous curvature have been also respectively
studied in \cite{TRrandom} and \cite{FCCrandom}.

The phase diagram of the model of \cite{npb95} has been first studied
in \cite{nostro3d} and \cite{BGGM} by the cluster variation method
(CVM) \cite{CVMrev}.  A sequence of transitions from the flat to the
piled--up phase through the partially folded octahedral and
tetrahedral phases was found by varying the bending rigidity from
$\infty$ to $-\infty$. The density--matrix renormalization group
(DMRG) calculations of \cite{Nishi1,Nishi2} have confirmed the
occurrence of these transitions. Both CVM and DMRG have predicted
first--order flat--octahedral and octahedral--tetrahedral transitions
while the character of the third transition is controversial.  CVM
suggested a continuous tetrahedral piled--up phase transition at
variance with the weak first--order behavior predicted by DMRG.

In order to further analyze the character of the crumpling transitions
of the triangular network in the fcc lattice, a modified folding rule
allowing defects in the network has been introduced in
\cite{Nishi3}. Transfer--matrix calculations of \cite{Nishi3}
confirmed the order of the transitions given by DMRG.

In the present work we observe that the model of folding with defects
is interesting by itself. Relaxing the local constraints on the Ising
variables in the model of Ref.\ \cite{npb95} corresponds to accepting
folding configurations with bending and/or meeting points in the folding
lines, which can be obtained by allowing cuts between adjacent
triangles. Therefore the model with progressively relaxed constraints
can describe polymerized membranes with an increasing number
of defects in the connectivity rules. Here we will introduce such a
model and study its equilibrium properties by means of the CVM.

The paper is organized as follows. In the next section we will
introduce the model and describe the CVM approximation scheme used for
studying it. Our results for the phase diagram and the equilibrium
behavior of the most important quantities are reported in Sect. III.
Conclusions will complete the paper.

\section{The model and the method}
\label{s:modello}
\par\noindent

The model for the folding of the
triangular lattice in the three--dimensional  fcc lattice has been
introduced in \cite{npb95,nostro3d,BGGM}.

In this model, sites of the triangular lattice are mapped onto sites
of the fcc lattice, with the condition that two sites that are
nearest--neighbors (NNs) in the triangular lattice must remain NNs in
the fcc lattice. Let us consider two adjacent plaquettes of the
triangular lattice, and call $\theta$ the angle formed by their normal
vectors (defined in such a way that $\theta$ always vanishes in the
planar configuration).  It is easy to see that upon mapping onto the
fcc lattice, the two plaquettes will be in one of four relative
orientations, namely: (i) no fold, with $\theta = 0, \cos\theta = 1$;
(ii) octahedral fold, with $\cos\theta = 1/3$ (the two plaquettes
belong to the same octahedron in the fcc lattice); (iii) tetrahedral
fold, with $\cos\theta = -1/3$ (they belong to the same tetrahedron in
the fcc lattice); (iv) complete fold, with $\cos\theta = -1$ (they are
on top of each other). A fold between two adjacent plaquettes has an
energy cost, due to curvature, given by $-K \cos \theta$.

The model allows various representations \cite{npb95}, a particularly
convenient one for our purposes being defined in terms of 2 sets of
Ising variables. These variables are usually denoted by $\sigma_i$ and
$z_i$, the pair $(\sigma_i,z_i)$ being associated to a plaquette of
the triangular lattice or, equivalently, to a site $i$ of the dual
hexagonal lattice $\Lambda$. Sites of $\Lambda$ correspond to
plaquette centers in the triangular lattice, while edges of $\Lambda$
are perpendicular to the edges of the triangular lattice. In terms of
these Ising variables, the fold between two adjacent plaquettes 1 and
2 (corresponding to two NN sites in $\Lambda$) is specified by the
values $\sigma_1\sigma_2$ and $z_1 z_2$ according to
Tab.\ \ref{t:folds}, and the angle $\theta$ is given by
\begin{equation}
\cos\theta = \sigma_1\sigma_2 \frac{1 + 2 z_1 z_2}{3}\;.
\label{costheta}
\end{equation}
In the language of magnetism, we can say that different folds
correspond to different types of domain walls for our Ising variables.

\begin{table}[t]
\begin{tabular}{c|c|c|c}
\hline\hline
Type of fold & $\cos\theta$ & $\sigma_1\sigma_2$ & $z_1 z_2$
\\
\hline\hline
no fold & $+1$ & $+1$ & $+1$
\\
\hline
octahedral & $+1/3$ & $-1$ & $-1$
\\
\hline
tetrahedral & $-1/3$ & $+1$ & $-1$
\\
\hline
complete fold & $-1$ & $-1$ & $+1$
\\
\hline\hline
\end{tabular}
\caption{Type of folds and their representation in terms of Ising
  variables.
}
\label{t:folds}
\end{table}

However, not all configurations of our Ising variables are allowed. It
has been shown in \cite{npb95} that the variables $z_i$ and $\sigma_i$
have to satisfy two constraints, or folding rules, in order to
describe a proper folding configuration over the fcc lattice. Given an
hexagon $e$ in the set $E$ of all the elementary plaquettes in
$\Lambda$, let $1,\dots,6$ be its six sites ordered counterclockwise,
$\sigma_e = \{ \sigma_i, i = 1, \ldots 6 \}$ and $z_e = \{ z_i, i = 1,
\ldots 6 \}$. Set
\begin{equation}
L(\sigma_e) = \sum_{i=1}^6 \sigma_i \label{rule1lhs}
\end{equation}
and
\begin{equation}
M_{c}(\sigma_e,z_e) = \sum_{i=1}^6 \frac{1 - z_i z_{i+1}}{2} \
\Delta_{i,c}(\sigma_e) \label{rule2lhs}
\end{equation}
where $z_7=z_1$ and
\begin{equation}
\Delta_{i,c}(\sigma_e)=\left\{
\begin{array}{ll}
1& \;\;\; {\rm if}\; \sum_{j=1}^i \sigma_j =  c\  {\rm mod}\ 3 \\
0& \;\;\; {\rm otherwise}
\end{array}
\right.
\end{equation}
with $i=1,\dots,6$ and $c=1,2$.
The folding rules then read
\begin{eqnarray}
&& L(\sigma_e) = 0 \ {\rm mod}\  3 \label{rule1}
\\
&& M_{c}(\sigma_e,z_e) = 0 \ {\rm mod}\ 2, \qquad c=1,2,3.
\label{rule2}
\end{eqnarray}
Notice that $M_1(\sigma_e,z_e) + M_2(\sigma_e,z_e) +
  M_3(\sigma_e,z_e)$ is an even number, and therefore the number of
  violated constraints in \ref{rule2} can be only 0 or 2: as a
  consequence, it is sufficient to impose only 2 of the 3 conditions in
  \ref{rule2}.


We now introduce defects in the
configurations of the triangular lattice. A violation of the
folding rules Eqs.(\ref{rule1}), (\ref{rule2})  at a hexagon
represents a defect at that hexagon.

The folding rules Eqs.(\ref{rule1}) and (\ref{rule2}) ensure that
after a complete turn around  any set of six triangles with a common
vertex the triangles occupy the same absolute position \cite{npb95}.
Clearly, defects alter the local connectivity of the original
triangular lattice and can be associated to cuts in the lattice.
More specifically, defects can be associated to certain kinds of
bending and/or meeting points of
the various domain  walls for the $(\sigma,z)$ variables, which are
not allowed in the original folding model and therefore do not
appear in the set of allowed vertices in Fig.\ 10 of \cite{npb95}.

\begin{figure*}[t]
\resizebox{24cm}{!}{\rotatebox{0}{\includegraphics{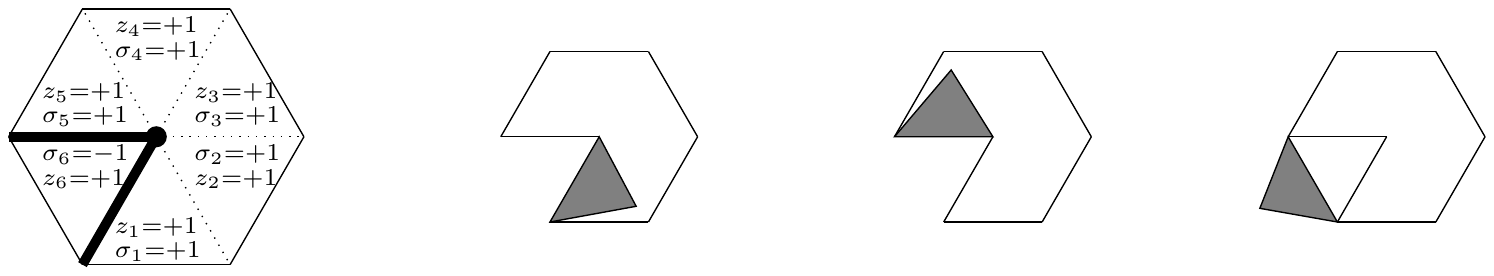}}}
 \vskip -10 cm
 \caption{On the left: a hexagon configuration with a complete fold
 domain wall represented by the 
 thick line between $(\sigma,z) = (+1,+1)$ and $(-1,+1)$. On the right:
the three possible surface folded configurations corresponding to
the hexagon spin configuration on the left, obtainable by cuts in
the original triangular network as described in  the main text.
}
 \label{f:rap1}
\end{figure*}

In Fig.~\ref{f:rap1},
for example, we have a
hexagon configuration containing only
a domain wall between $(\sigma,z) = (+1,+1)$ and $(-1,+1)$ (we call
it a complete--fold domain wall),
which shows a bend at the center of the hexagon. This corresponds to
a violation of constraint (\ref{rule1}) only, as one can easily
check. This configuration cannot appear in a fold of the triangular
lattice, unless one allows cuts. It is however immediately evident
that, at least locally, there are three possible ways to cut the
hexagon. One can select one of the three edges of triangle number 6,
cut the remaining two edges, and then fold the triangle along the
selected edge. If this were the only defect in the whole
configuration, this argument would apply globally as well.
Therefore, already on the basis of this example, it is clear that a
unique reconstruction of the surface is not feasible in the presence
of defects.

\begin{figure*}[t]
\resizebox{24cm}{!}{\rotatebox{0}{\includegraphics{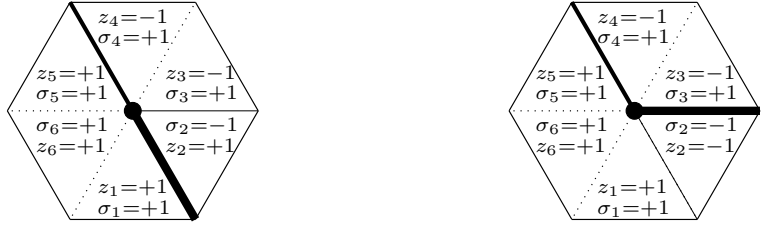}}}
 \vskip -10 cm
 \caption{Two different hexagon configurations where  a complete fold,
  a tetrahedral and an octahedral domain wall (respectively represented by lines of
decreasing thickness) meet at the center of the hexagon. The
octahedral and the complete fold domain walls are swapped in the two
configurations.       }
 \label{f:rap2}
\end{figure*}

The analysis of a similar
configuration with
$(\sigma_i,z_i) = (+1,+1), i=1,  \ldots, 5$ and
$(\sigma_6,z_6)=(-1,-1)$, representing the bending of an
octahedral--fold
domain wall
 shows that in this case both
constraints (2.5) and (2.6) are violated. On the other hand, in the
case of bending of a tetrahedral--fold
domain wall
between $(\sigma,z) = (+1,+1)$ and $(+1,-1)$,
only (\ref{rule2}) would be violated. On the basis of the previous
examples one might be tempted to consider  a specific geometrical
interpretation for the separate violation of each of the two folding
rules. This other example, however, shows that this interpretation
would be problematic. In Fig.~\ref{f:rap2}, we see a configuration
where an octahedral--fold
domain wall  meets (ending at the meeting point) another domain
wall,
 which at the meeting point changes its character from
tetrahedral to complete: in this case both constraints (2.5) and
(2.6) are violated. If, however, the complete and octahedral
domain walls
are swapped, only (2.5) is violated.
One can conclude from the above examples that the two folding rules
can not be directly associated to definite classes of defects.

Due to the above observation, we do
not find reasons to attribute different energy costs to different
kinds of defects and will consider a model where all defects are
weighted in the same way \cite{Nota}. A single energy parameter
$\lambda$ will be coupled to the number of defects, that is the
number of hexagons at which (one or more) constraints are violated.
Taking also  into  account  Eq.\ (\ref{costheta}) for  writing the
curvature energy as in \cite{nostro3d}, we are led to consider the
Hamiltonian (energies are given in units of $k_{\rm B}T$)
\begin{equation}
\label{hamilton}
\begin{array}{rcl}
\!\!\!
H
&\!\!=&\!\!
{\displaystyle
 -\frac{K}{3}
  \sum_{\langle i j \rangle} \sigma_i \sigma_j (1 + 2 z_i z_j)
 \vphantom{\bigg\{_\Big\{}
}
\\
\!\!\!
&&\!\!
{\displaystyle
 -\lambda \sum_{e\in E}
 \Big[
      \fcar_{\{L(\sigma_e)=0\ \textrm{mod}\ 3\}}
}
\\
\!\!\!
&&\!\!
{\displaystyle
 \phantom{
          +\sum_{e\in E}
          \Big[
         }
      \times
      \prod_{c=1}^3\fcar_{\{M_{c}(\sigma_e,z_e)=0\ \textrm{mod}\ 2\}}
 \Big]
}
\end{array}
\end{equation}
where $\fcar_{\{\textrm{condition}\}}$ is equal to one if the
condition is satisfied and to zero otherwise.  Note that in the above
formula the first sum is extended to the NN pairs and the second to
the hexagons of the lattice.  Moreover, the product over the
  three color indices in the last sum is redundant: according to our
  discussion of constraints (\ref{rule2}), it is sufficient to impose
  only 2 of the 3 constraints.

Finally, we observe that it is not
  possible to express the energy cost of defects in terms of local
  weights for the length of the cuts needed to obtain a surface realization of a
given $(\sigma,z)$ configuration. This can be shown by the following
example.
Consider a set of 3 defects corresponding
 to
$60^\circ$ bends of complete--fold domain walls as those shown in
Fig.~\ref{f:rap1}.
These defects can be used to construct configurations where a
complete--fold domain wall  can form equilateral triangles of any
size, $(\sigma,z) = (+1,+1)$ outside the triangle, and $(\sigma,z) =
(-1,+1)$ inside it. Hence, given the same set of 3 defects, one can
construct configurations corresponding to cuts of any size.


 As it was done for the two--dimensional model discussed in
\cite{nostro} and for the defect free three--dimensional case studied
in \cite{nostro3d}, the phase diagram will be investigated by means
of the hexagon approximation of the CVM. For the sake of
self--containedness we briefly recall the main features of our
approach.

We recall that $E$ is the collection of hexagons on the lattice. We
denote by $M$ the collections of all the hexagons and all their
sub--clusters (site subsets). To each cluster $\alpha\in M$ a
probability distribution $\rho_\alpha(\sigma_\alpha,z_\alpha)$ is
associated, where $\sigma_\alpha = \{ \sigma_i, i \in \alpha \}$ and
$z_\alpha = \{ z_i, i \in \alpha \}$.  In this CVM approximation the
free energy functional is given by
\begin{equation}
\label{cvmfunc}
\begin{array}{rcl}
F_M
&\!\!=&\!\!
{\displaystyle
  \sum_{\alpha\in M}
  \sum_{\sigma_\alpha,z_\alpha}H_\alpha(\sigma_\alpha,z_\alpha)
                             \rho_\alpha(\sigma_\alpha,z_\alpha)
  \vphantom{\bigg\{_\Big\}}
}
\\
&&\!\!
{\displaystyle
  +
  \sum_{\alpha\in M}
  a_\alpha
  \sum_{\sigma_\alpha,z_\alpha}
  \rho_\alpha(\sigma_\alpha,z_\alpha)
  \log\rho_\alpha(\sigma_\alpha,z_\alpha)
},
\\
\end{array}
\end{equation}
where the coefficients $a_\alpha$ are such that
for each $\alpha\in M$
\begin{equation}
\label{an}
\sum_{\beta\in M:\,\beta\supseteq\alpha}a_\beta=1
\end{equation}
and the Hamiltonian terms $H_\alpha$ are defined as follows:
for a cluster made of two neighboring sites $\langle i,j\rangle$
we have
\begin{displaymath}
H_{\langle i,j\rangle}
=
 -\frac{K}{3}
  \sigma_i \sigma_j (1 + 2 z_i z_j),
\end{displaymath}
for a cluster made of a hexagon $e\in E$
\begin{displaymath}
H_e = - \lambda\fcar_{\{L(\sigma_e)=0\ \textrm{mod}\ 3\}}
\prod_{c=1}^2\fcar_{\{M_{c}(\sigma_e,z_e)=0\ \textrm{mod}\ 2\}}
\end{displaymath}
and $H_\alpha = 0$ otherwise.  By using Eq.\ (\ref{an}) it is not difficult
to prove that the sole non--vanishing coefficients $a_\alpha$ are those
associated to hexagons, NN pairs, and single sites;
coefficients associated to five--site, four--site, three--site, and
not neighboring two--site clusters are all equal to zero. In
particular we have that
\begin{displaymath}
a_e=1,\;\;
a_{\langle i,j\rangle}=-1,
\;\textrm{ and }\;
a_i=1
\end{displaymath}
for each hexagon $e\in E$,
each nearest--neighbor pair $\langle i,j\rangle$,
and each site $i\in\Lambda$.

We now exploit the translational invariance of the system, which implies
that all the probability distributions associated to a particular
sub--family of clusters in $M$ are equal. Thus, we denote by
$\rho_6(\sigma_1,\dots,\sigma_6,z_1,\dots,z_6)$ the probability
distribution associated to hexagons and by
$\rho_2(\sigma_1,\sigma_2,z_1,z_2)$ that associated to NN pairs.  Site
clusters form two sub--families: indeed, a hexagonal lattice is a
bipartite lattice, made of two inter--penetrating (triangular)
sub--lattices $a$ and $b$, such that all the NNs of a site in $a$ belong
to $b$ and vice-versa. Since we expect the symmetry between the two
sub--lattices to be broken in some thermodynamic phases of the model, it
is important to distinguish the site probability distributions
corresponding to $a$ and $b$. For the same reason, in the argument of
$\rho_2$, the first and the third entries ($\sigma_1$ and $z_1$) refer
to sub--lattice $a$, while the second and the fourth ones ($\sigma_2$
and $z_2$) refer to sub--lattice $b$, and they cannot be interchanged.

With the above definitions and notations, and observing that
the number of hexagons and NN pairs in $\Lambda$ are respectively
$N_6 = |\Lambda|/2$ and $N_2 = 3|\Lambda|/2$, from Eq.\ (\ref{cvmfunc}) we
obtain the following expression for the CVM free energy density
functional $f=F_M/|\Lambda|$:
\begin{widetext}
\begin{eqnarray}
\label{cvmfree}
f
&=&
{\displaystyle
 -\frac{1}{2}K
 \textrm{Tr}_2[\sigma_1 \sigma_2 (1 + 2 z_1 z_2) \rho_2]
 -\frac{1}{2}
 \textrm{Tr}_6
 \Big[
 \Big(\lambda\fcar_{\{L(\sigma_e)=0\ \textrm{mod}\ 3\}}
 \prod_{c=1}^2
 \fcar_{\{M_{c}(\sigma_e,z_e)=0\ \textrm{mod}\ 2\}}\Big)\rho_6
 \Big]
 \vphantom{\bigg\{_\big\}}
}
\\
&&
{\displaystyle
 + \frac{1}{2} \textrm{Tr}_6[\rho_6 \ln \rho_6]
 - \frac{3}{2} \textrm{Tr}_2[\rho_2 \ln \rho_2]
 +\frac{1}{2} \textrm{Tr}_1[\rho_{1,a} \ln \rho_{1,a}]
 +\frac{1}{2} \textrm{Tr}_1[\rho_{1,b} \ln \rho_{1,b}]
 + \nu(\textrm{Tr}_6 \rho_6 - 1),
}
\nonumber
\end{eqnarray}
\end{widetext}
where we have introduced the notation
\begin{displaymath}
\textrm{Tr}_6
=
\sum_{\newatop{\sigma_1,\dots,\sigma_6}{z_1,\dots,z_6}},\;\;
\textrm{Tr}_2
=
\sum_{\newatop{\sigma_1,\sigma_2}{z_1,z_2}},
\textrm{Tr}_1
=
\sum_{\sigma_1,z_1},
\end{displaymath}
and
$\nu$ is a Lagrange multiplier which ensures the normalization of
$\rho_6$. The other probability distributions do not need a
normalization constraint since they can be written as partial traces
of probability distributions of larger clusters. More precisely,
\begin{widetext}
\begin{equation}
\label{tracce}
\begin{array}{l}
\rho_2(\sigma_1,\sigma_2,z_1,z_2)
=
 \vphantom{\bigg\{_\big\}}
\\
\phantom{aa}
{\displaystyle
 \frac{1}{6}
 \sum_{\newatop{\sigma_3,\dots,\sigma_6}{z_3,\dots,z_6}}
 [
 \rho_6(\sigma_1,\sigma_2,\sigma_3,\sigma_4,\sigma_5,\sigma_6,\dots)
 +\rho_6(\sigma_3,\sigma_2,\sigma_1,\sigma_4,\sigma_5,\sigma_6,\dots)
 +\rho_6(\sigma_3,\sigma_4,\sigma_1,\sigma_2,\sigma_5,\sigma_6,\dots)
}
\\
\phantom{aaaaaaaaaaa}
{\displaystyle
 +\rho_6(\sigma_3,\sigma_4,\sigma_5,\sigma_2,\sigma_1,\sigma_6,\dots)
 +\rho_6(\sigma_3,\sigma_4,\sigma_5,\sigma_6,\sigma_1,\sigma_2,\dots)
 +\rho_6(\sigma_1,\sigma_4,\sigma_5,\sigma_6,\sigma_3,\sigma_2,\dots)
 ]
}
\\
\end{array}
\end{equation}
\end{widetext}
where the $z$ variables in the argument of $\rho_6$ appear in the same
order as $\sigma$ variables, and
\begin{eqnarray}
\label{tracce2}
\rho_{1,a}(\sigma_1,z_1)
&=&
 \sum_{\sigma_2,z_2}
 \rho_2(\sigma_1,\sigma_2,z_1,z_2),
 \vphantom{\bigg\{_\Big\}} \\
\rho_{1,b}(\sigma_2,z_2)
&=&
 \sum_{\sigma_1,z_1}
 \rho_2(\sigma_1,\sigma_2,z_1,z_2).
 \vphantom{\bigg\{_\Big\}}
\end{eqnarray}

With the above definitions the CVM free energy density functional $f$
can be regarded as a function of $\rho_6$ only.  The minimization must
be performed numerically, and this can be easily done by standard
iterative methods as in \cite{nostro} (see \cite{CVMrev} for a survey
of such algorithms).
The
simplest possibility is to write stationarity equations by taking
derivatives of $f$ with respect to an element of
$\rho_6(\sigma_1,\dots,\sigma_6,z_1,\dots,z_6)$, for
some generic choice
$\sigma_1,\dots,\sigma_6,z_1,\dots,z_6=\pm1$ of the spin variables and letting
$\sigma_7=\sigma_1$ and $z_7=z_1$. After some
algebra, we get
\begin{widetext}
\begin{equation}
\label{staz}
\begin{array}{l}
\!\!\!
\!\!\!
\rho_6(\sigma_1,\dots,\sigma_6,z_1,\dots,z_6)
\\
{\displaystyle
=
\exp\{(K/6)\sum_{i=1}^6\sigma_i\sigma_{i+1}(1+2z_iz_{i+1})
 -2\nu
 +\lambda\fcar_{\{L(\sigma_e)=0\ \textrm{mod}\ 3\}}
  \prod_{c=1}^2\fcar_{\{M_{c}(\sigma_e,z_e)=0\ \textrm{mod}\ 2\}}\}
  \vphantom{\bigg\{_\Big\}}
}
\\
\phantom{a}
\phantom{=\,}
\times
[\rho_2(\sigma_1,\sigma_2,z_1,z_2)
 \rho_2(\sigma_3,\sigma_2,z_3,z_2)
 \rho_2(\sigma_3,\sigma_4,z_3,z_4)
 \rho_2(\sigma_5,\sigma_4,z_5,z_4)
 \rho_2(\sigma_5,\sigma_6,z_5,z_6)
 \rho_2(\sigma_1,\sigma_6,z_1,z_6)]^{1/2}
\\
\phantom{a}
\phantom{=\,}
\times
[\rho_{1,a}(\sigma_1,z_1)
 \rho_{1,b}(\sigma_2,z_2)
 \rho_{1,a}(\sigma_3,z_3)
 \rho_{1,b}(\sigma_4,z_4)
 \rho_{1,a}(\sigma_5,z_5)
 \rho_{1,b}(\sigma_6,z_6)
]^{-1/3}
\\
\end{array}
\end{equation}
\end{widetext}
that can be solved numerically with an iterative
approach.

\section{Results}
\label{s:risultati}
\par\noindent

We now present our results for the phase diagram of the folding model
Eq.\ (\ref{hamilton}), obtained by finding the stable (lowest free
energy) solutions of Eq.\ (\ref{staz}) at varying curvature energy and
defect cost.
 Following
\cite{BGGM} we introduce the order parameters $\parametro{O}\equiv
\langle \sigma^{\rr{st}}_i\rangle$, $\parametro{T}\equiv \langle
z_i\sigma^{\rr{st}}_i\rangle$ and $\parametro{P}\equiv \langle
\sigma^{\rr{st}}_i\rangle$. $\parametro{O}$ and $\parametro{T}$
  were named octahedral and tetrahedral order parameters,
   respectively, and indeed they are the only order parameters which
  become non--zero in the corresponding phases (observe, as explained later,
  that the octahedral phase of the model without defects
  will be called in this paper
  p--octahedral).
  $\parametro{P}$ was
  named planar order parameter, since it becomes non--zero in the
  planar (here called flat) phase. In order to characterize and
  distinguish all the phases we obtain here, it is convenient to
  define two additional order parameters, $\parametro{M}\equiv \langle
  \sigma_i\rangle$ and $\parametro{N}\equiv \langle z_i
  \sigma_i\rangle$. $\parametro{M}=1$ corresponds to configurations
  with tetrahedral folds or no folds only, while $\parametro{N}=1$ corresponds to configurations
  with octahedral folds or no folds only.
$\parametro{M}$ is non--zero only in the flat
  phase, while $\parametro{N}$ is non--zero in the flat and in the  f--octahedral
  phase, as described  later.
  The identification of the various phases of the model in
terms of the above order parameters is summarized in
Tab.~\ref{t:fasi}.

\begin{table}[t]
\begin{tabular}{c|c|c|c|c|c}
\hline\hline
 Phase & \parametro{M} & \parametro{P} & \parametro{N} & \parametro{O} &
         \parametro{T}
\\
\hline\hline
flat  & $\neq0$ & $\neq0$ & $\neq0$ & $0$ & $0$ \\
\hline
p--octahedral  & $0$ & $0$ & $0$ & $\neq0$ & $0$
\\
\hline
tetrahedral  & $0$ & $0$ & $0$ & $0$ & $\neq0$
\\
\hline
piled--up  & $0$ & $\neq0$ & $0$ & $\neq0$ & $\neq0$
\\
\hline
f--octahedral & $0$ & $0$ & $\neq0$ & $0$ & $0$
\\
\hline
disordered & $0$ & $0$ & $0$ & $0$ & $0$
\\
\hline\hline
\end{tabular}
\caption{Phases. In the first column the name of the phase is
reported.
In the columns from the second to the sixth it is indicated which, among the
order parameters
$\parametro{M}=\langle\sigma_i\rangle$,
$\parametro{P}=\langle z_i\rangle$,
$\parametro{O}=\langle\sigma^\rr{st}_i\rangle$,
$\parametro{N}=\langle\sigma_i z_i\rangle$,
and
$\parametro{T}=\langle\sigma^\rr{st}_iz_i\rangle$,
differ from zero.
}
\label{t:fasi}
\end{table}

\begin{figure*}
\begin{picture}(300,400)
\put(-120,0)
{
\resizebox{18cm}{!}{\rotatebox{0}{\includegraphics{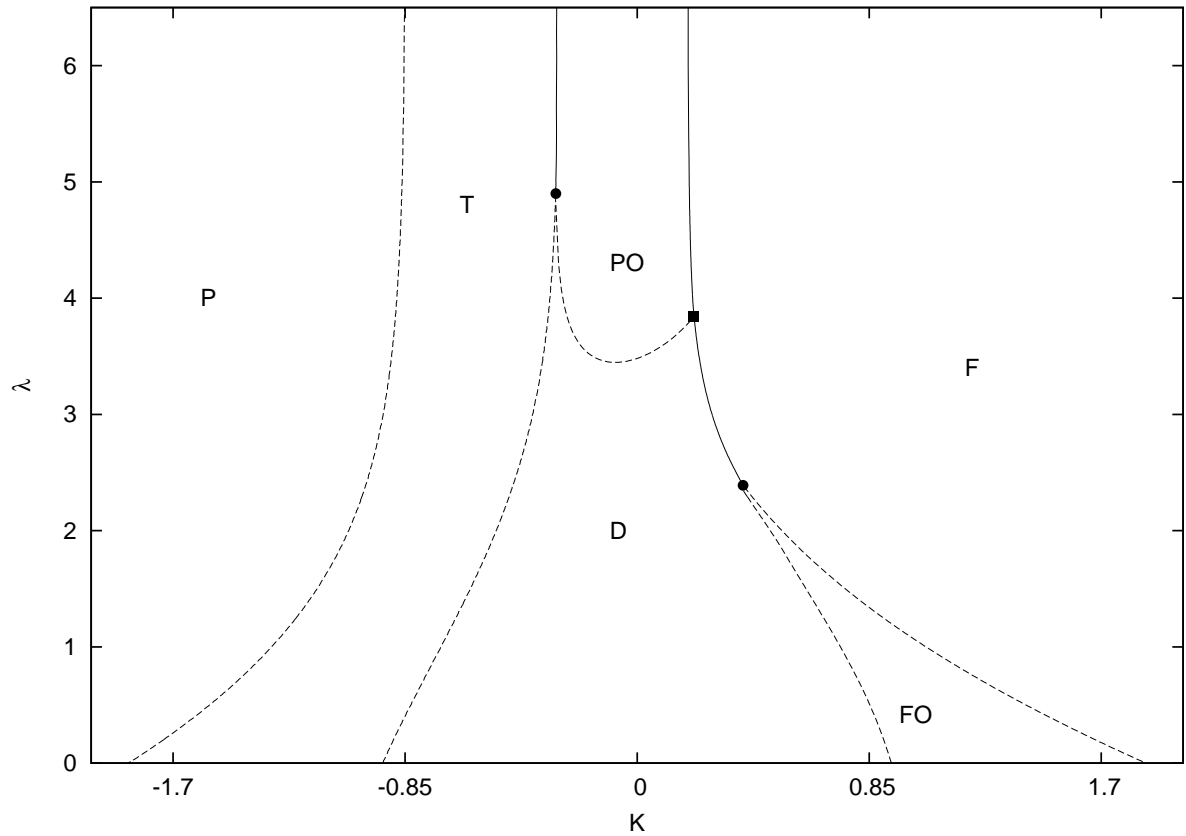}}}
}
\end{picture}
\caption{Phase diagram of the model Eq.\ (\ref{hamilton}), where
P, T, PO, D, FO, and F
stand respectively for piled--up, tetrahedral, piled--octahedral,
disordered, flat--octahedral, and flat.
The solid circles and the solid box
denote,
respectively, the bicritical points and  the critical end--point.
Solid and dashed lines denote, respectively, first--order and continuous
transitions.
}
\label{f:fase}
\end{figure*}

Fig.~\ref{f:fase} summarizes our results for the phase diagram in the
plane $K$--$\lambda$. We shall describe it by first considering the
large $\lambda$ limit, where defects are absent and one recovers
previous results \cite{nostro3d,BGGM}. We shall then consider smaller
values of $\lambda$ in order to see how the various phases and phase
transitions are modified by the introduction of defects.

At large enough $\lambda$, folding rules are practically never
violated and the equilibrium thermodynamics of the model becomes
independent of $\lambda$. Indeed, in the upper portion of the phase
diagram Fig.~\ref{f:fase}, the phase transition lines are
practically vertical and do not change their nature anymore.  We can
easily check that the results of the defect--free case
\cite{nostro3d,BGGM} are recovered in this limit. For instance, at
$\lambda=20$ and vanishing $K$  we obtain the entropy (per site) $S
= \ln q$, where $q=1.42805$, equal to the value obtained by the
constrained CVM approach of \cite{nostro3d} and in very good
agreement with the transfer matrix estimate $q=1.43(1)$ of
\cite{npb95}. Furthermore we obtain $\parametro{O}=0.87456$,
 $\parametro{M}=\parametro{P}=\parametro{N}=\parametro{T}=0$,
indicating a marked preference of the triangular lattice for
wrapping on an octahedron at zero curvature cost. As already noticed
in \cite{BGGM}, here most folds between adjacent plaquettes are
octahedral or complete. The sequence of folding transitions observed
in the defect--free limit \cite{nostro3d,BGGM} is reproduced here
already at $\lambda \gtrsim 4.91$.  Choosing $\lambda = 10$ as an
example, we find a flat phase at large $K$ (in the limit $\lambda
\to \infty$ this phase is perfectly flat in this approximation),
then at $K=0.1856$ a first--order transition occurs between the flat
and the octahedral phases, characterized by $\parametro{M}=0$,
$\parametro{P}=0$, $\parametro {N}=0$, $\parametro{O}=0.8247$, and
$\parametro{T}=0$. This phase, as shown in \cite{BGGM}, is also
characterized by a relative abundance of complete folds with respect
to other folds. In order to make a difference with another phase
appearing  in the phase diagram (see the following) that is
characterized by a relevant presence of octahedral folds, it will be
called p(iled-up)--octahedral phase. Upon further decreasing $K$ we
find another first--order transition at $K=-0.2940$ between the
p--octahedral phase with $\parametro{M}=0$, $\parametro{P}=0$,
$\parametro{N}=0$, $\parametro{O}=0.5816$, and $\parametro{T}=0$ and
the tetrahedral one with $\parametro{M}=0$, $\parametro{P}=0$,
$\parametro{N}=0$, $\parametro{O}=0$, and $\parametro{T}=0.7466$,
Finally, at $K=-0.8395$ there is a continuous transition from the
tetrahedral phase to the piled--up phase with the order parameters
characterized by the continuous vanishing of $\parametro{P}$ and
$\parametro{O}$, while $\parametro{T}=0.9993$. This sequence of
phases and phase transitions agrees in nature with that found in the
defect--free limit \cite{nostro3d,BGGM}, providing a confirmation of
the validity of the present approach.

\begin{figure*}
\begin{picture}(500,600)
\put(-20,0)
{
\resizebox{10cm}{!}{\rotatebox{0}{\includegraphics{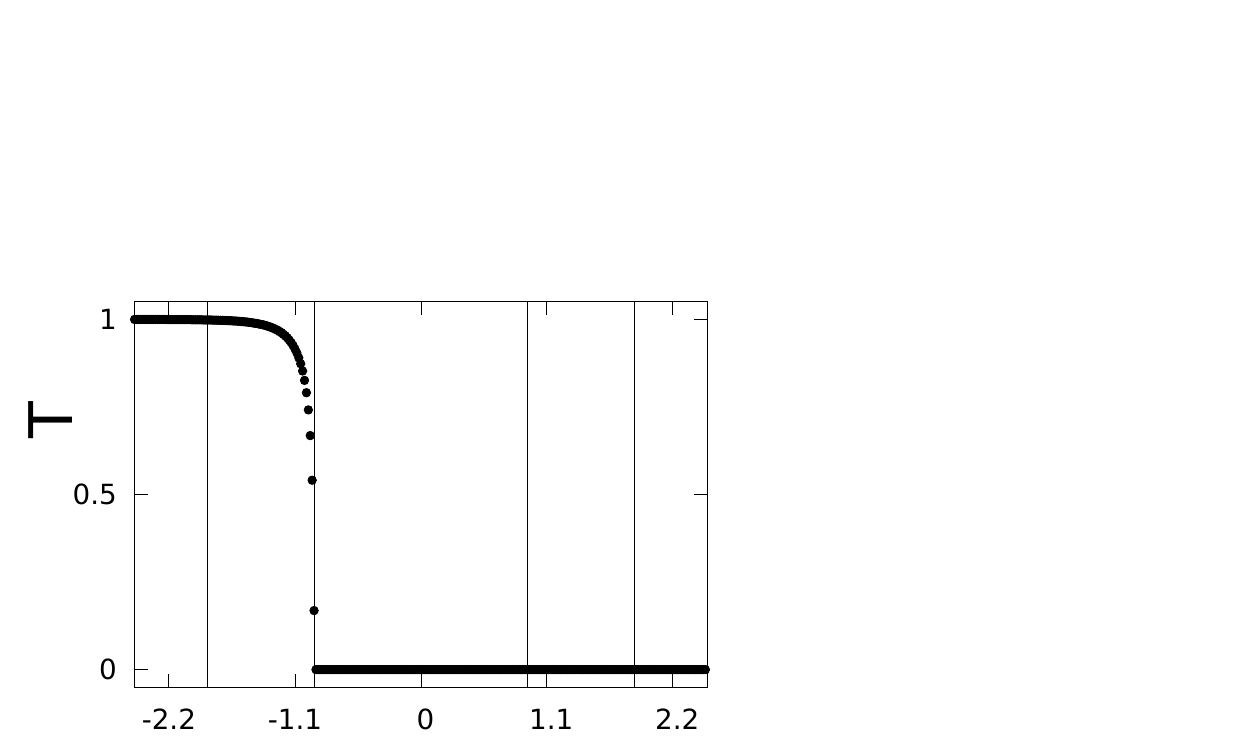}}}
}
\put(150,0)
{
\resizebox{10cm}{!}{\rotatebox{0}{\includegraphics{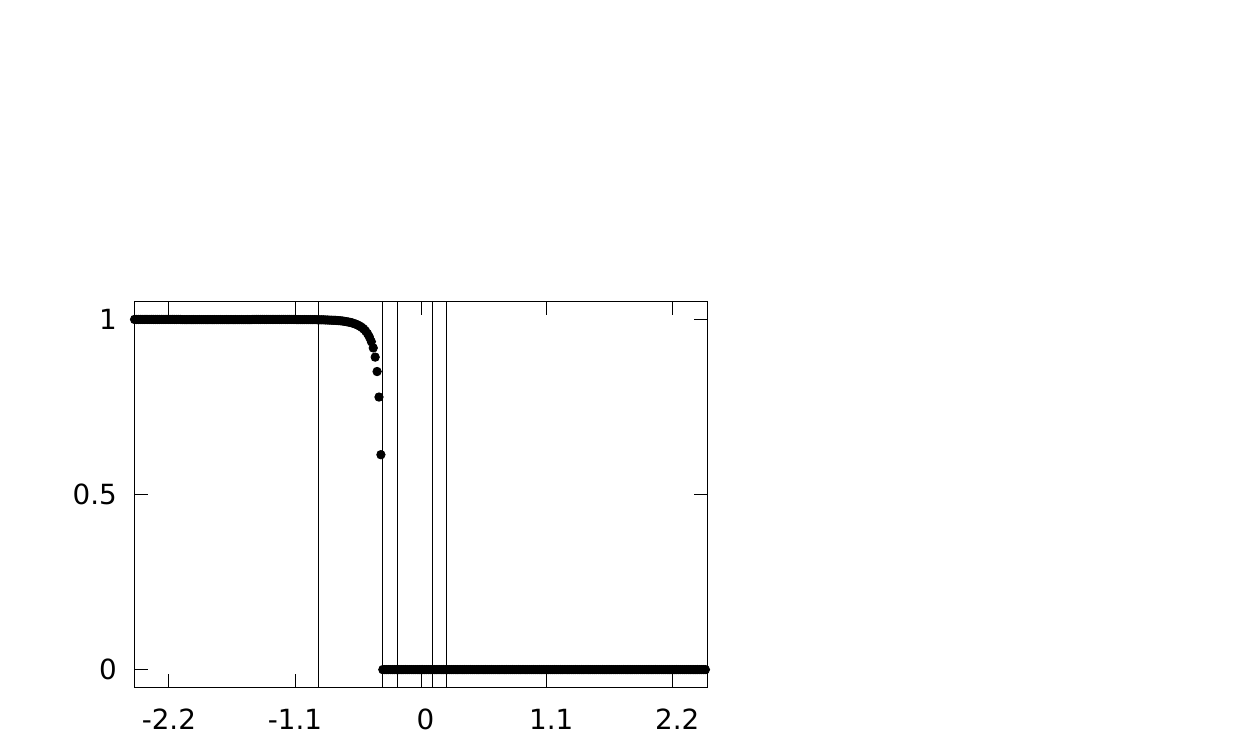}}}
}
\put(320,0)
{
\resizebox{10cm}{!}{\rotatebox{0}{\includegraphics{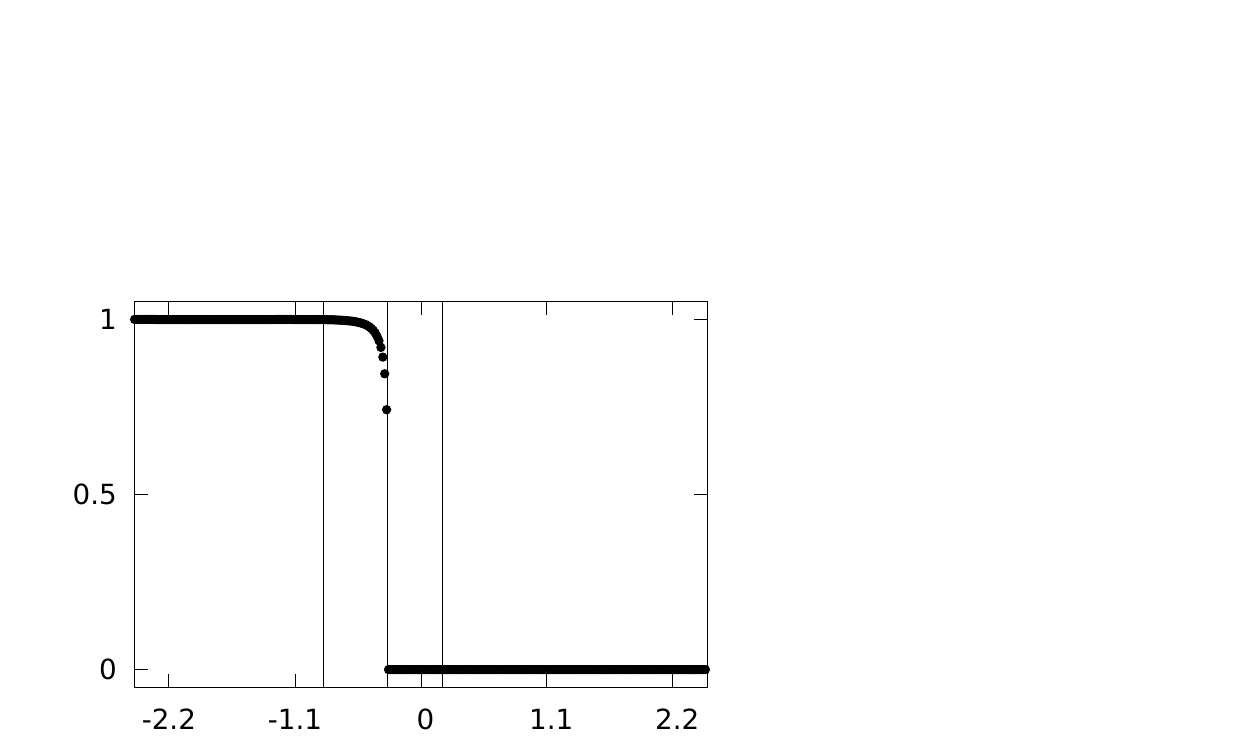}}}
}
\put(-20,115)
{
\resizebox{10cm}{!}{\rotatebox{0}{\includegraphics{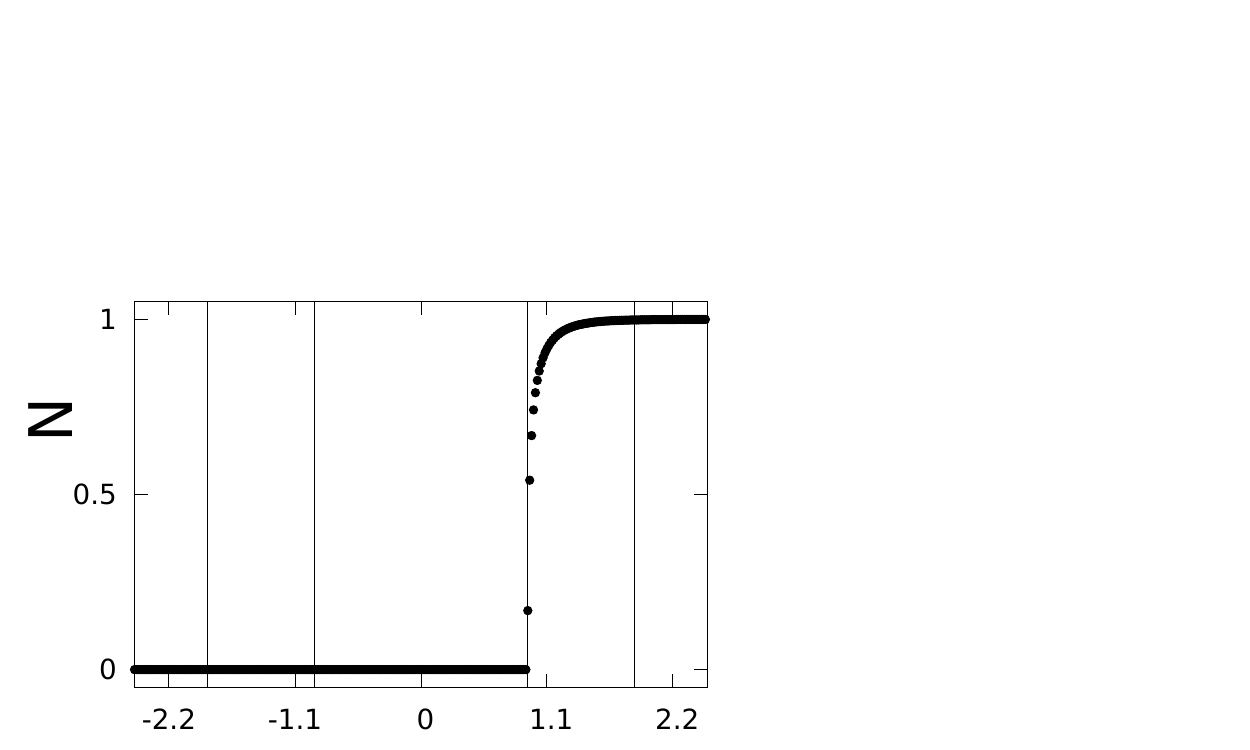}}}
}
\put(150,115)
{
\resizebox{10cm}{!}{\rotatebox{0}{\includegraphics{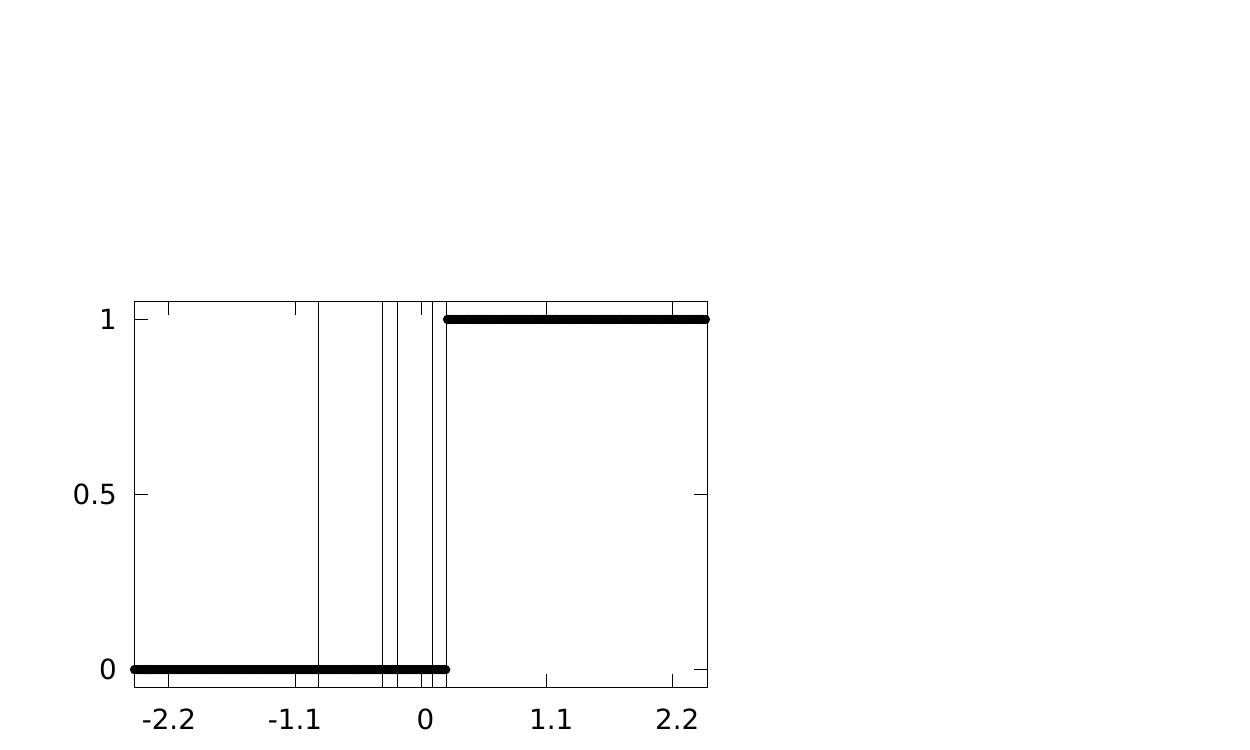}}}
}
\put(320,115)
{
\resizebox{10cm}{!}{\rotatebox{0}{\includegraphics{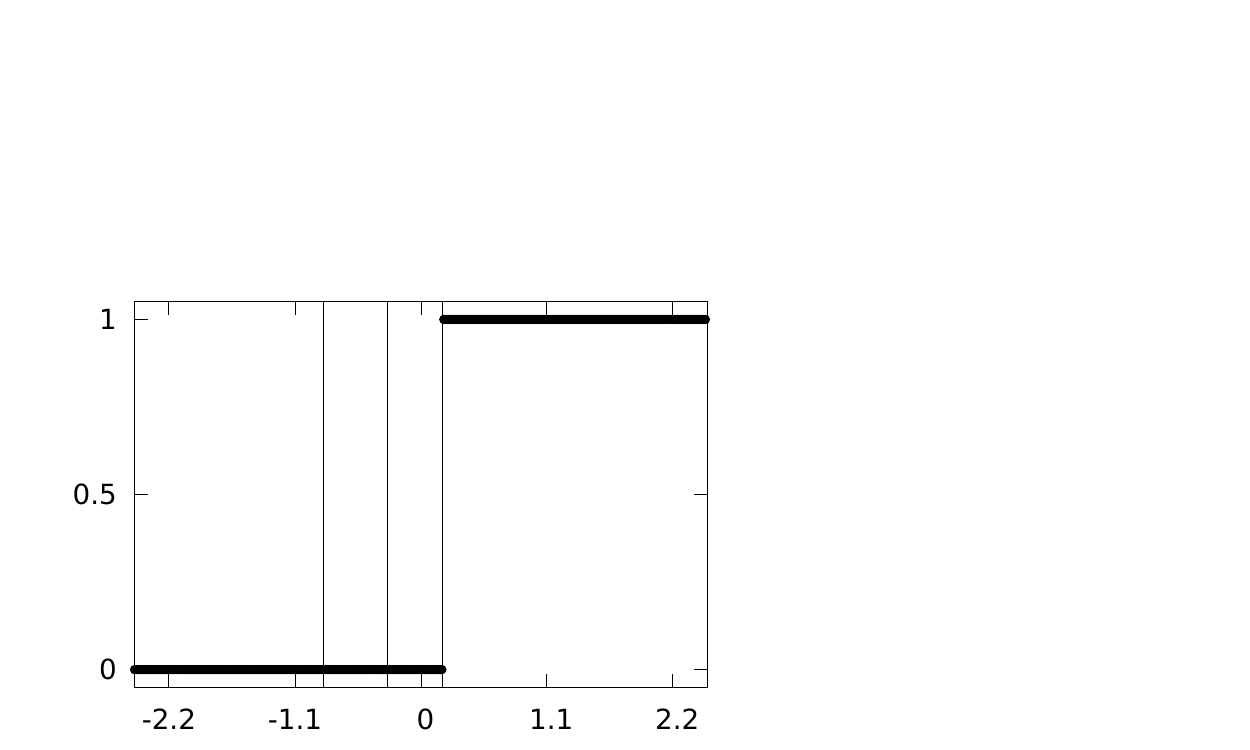}}}
}
\put(-20,230)
{
\resizebox{10cm}{!}{\rotatebox{0}{\includegraphics{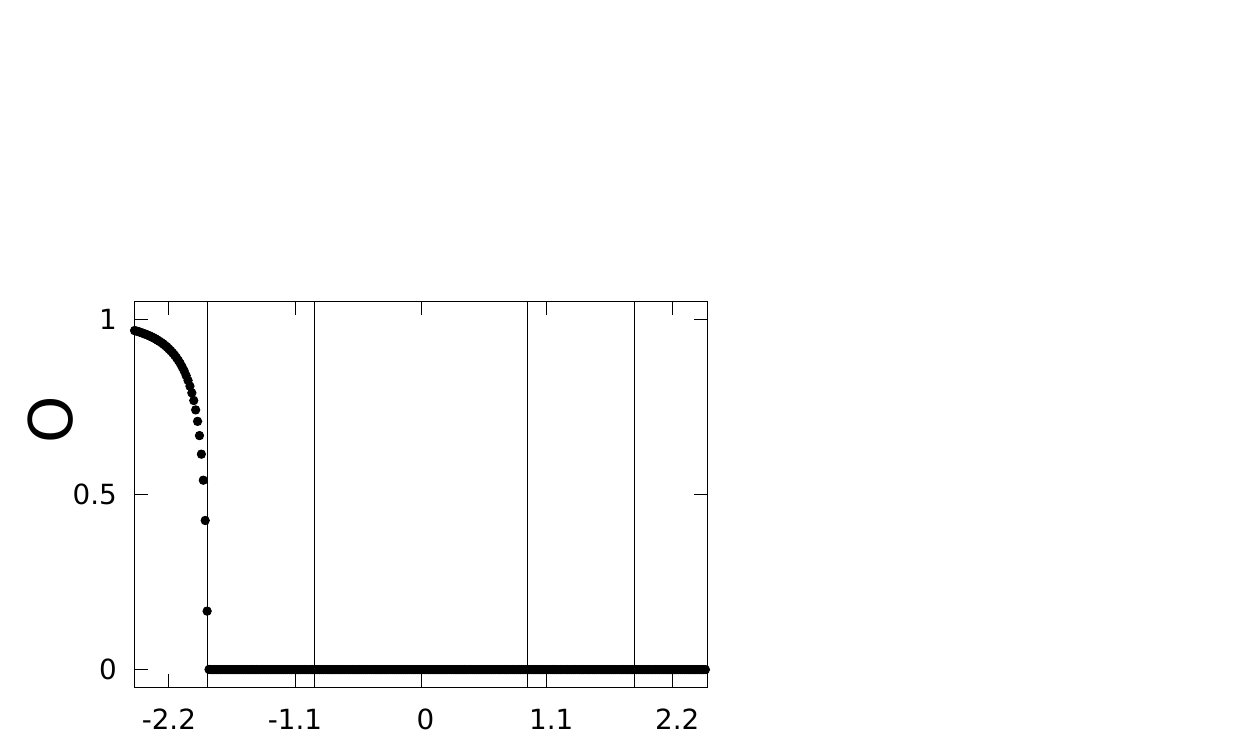}}}
}
\put(150,230)
{
\resizebox{10cm}{!}{\rotatebox{0}{\includegraphics{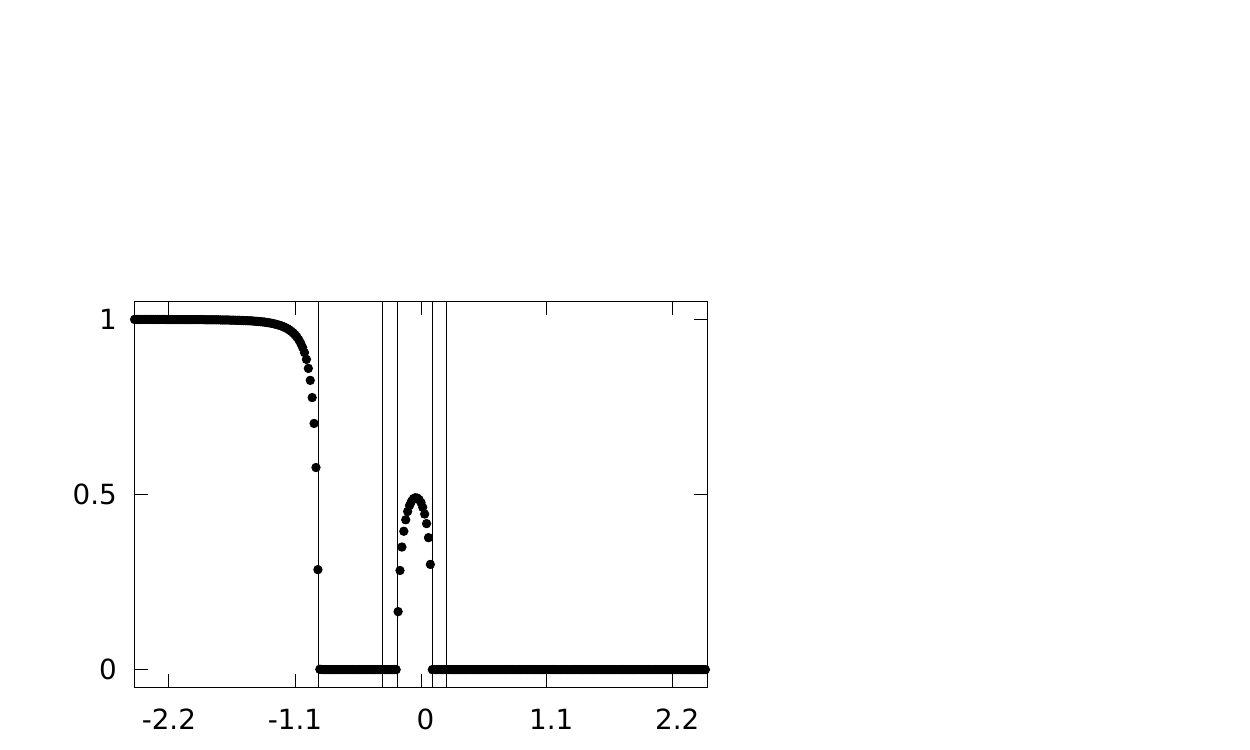}}}
}
\put(320,230)
{
\resizebox{10cm}{!}{\rotatebox{0}{\includegraphics{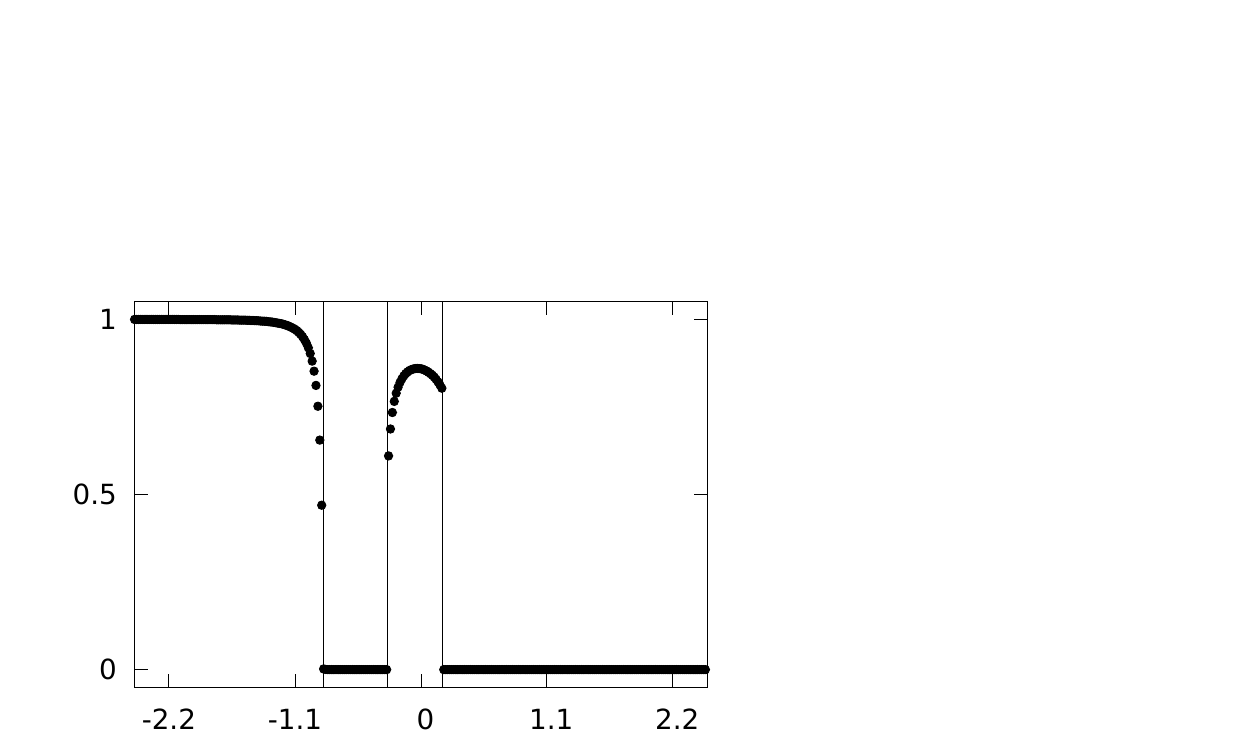}}}
}
\put(-20,345)
{
\resizebox{10cm}{!}{\rotatebox{0}{\includegraphics{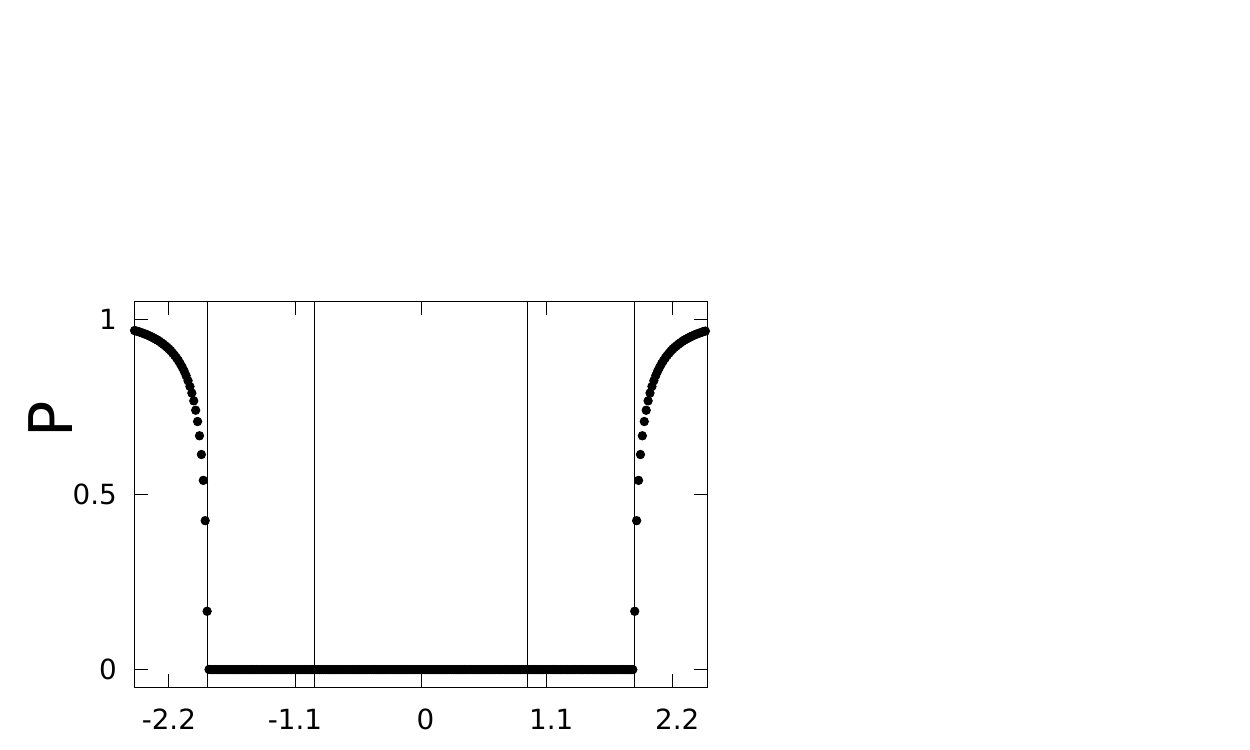}}}
}
\put(150,345)
{
\resizebox{10cm}{!}{\rotatebox{0}{\includegraphics{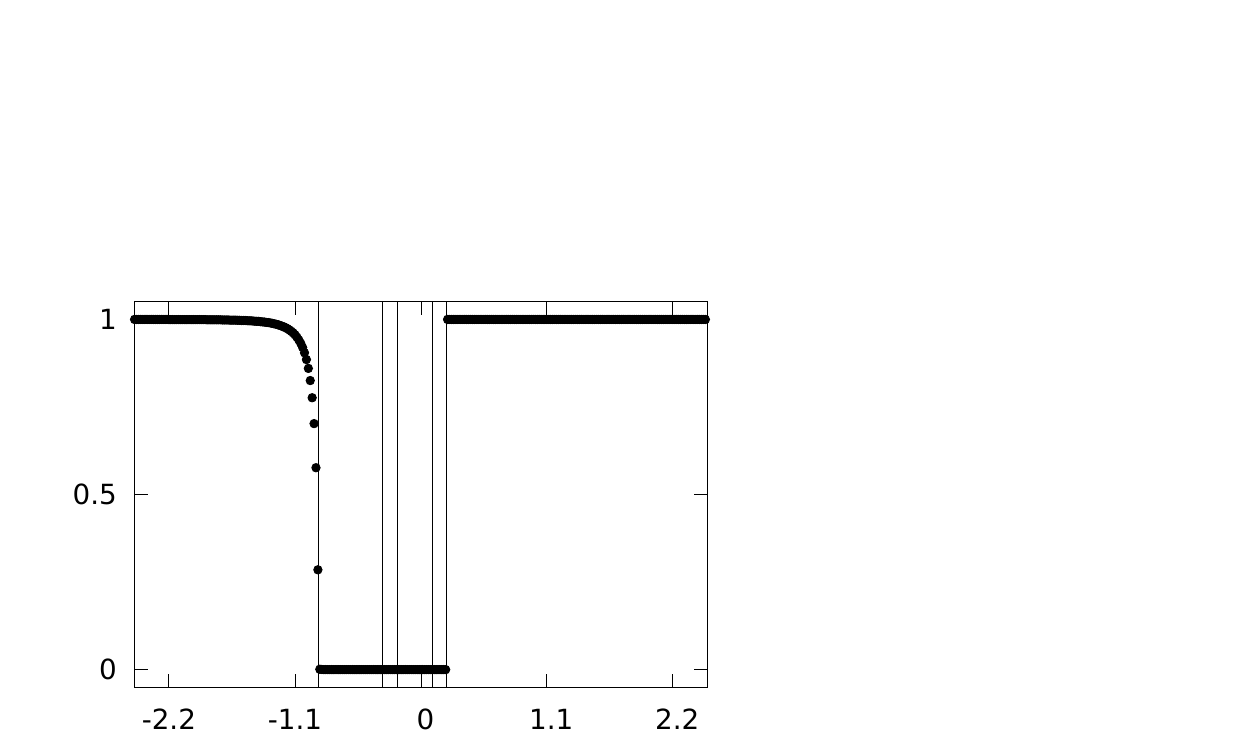}}}
}
\put(320,345)
{
\resizebox{10cm}{!}{\rotatebox{0}{\includegraphics{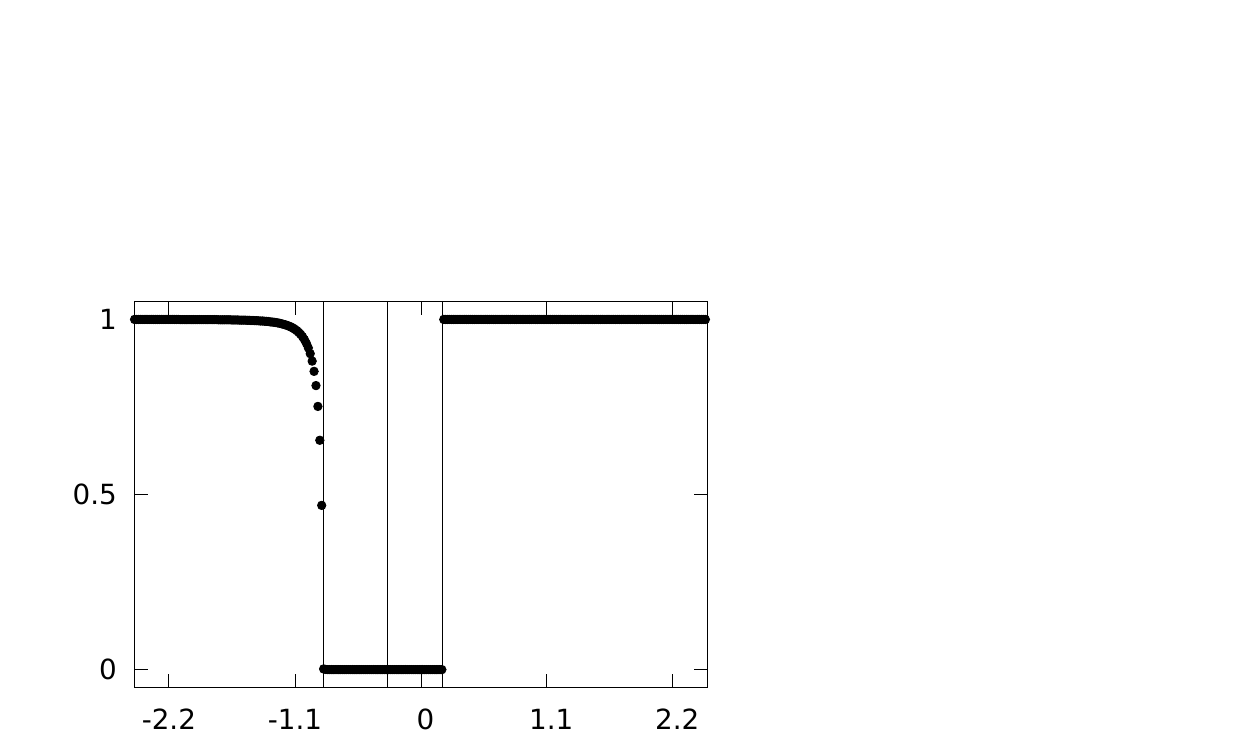}}}
}
\put(-20,460)
{
\resizebox{10cm}{!}{\rotatebox{0}{\includegraphics{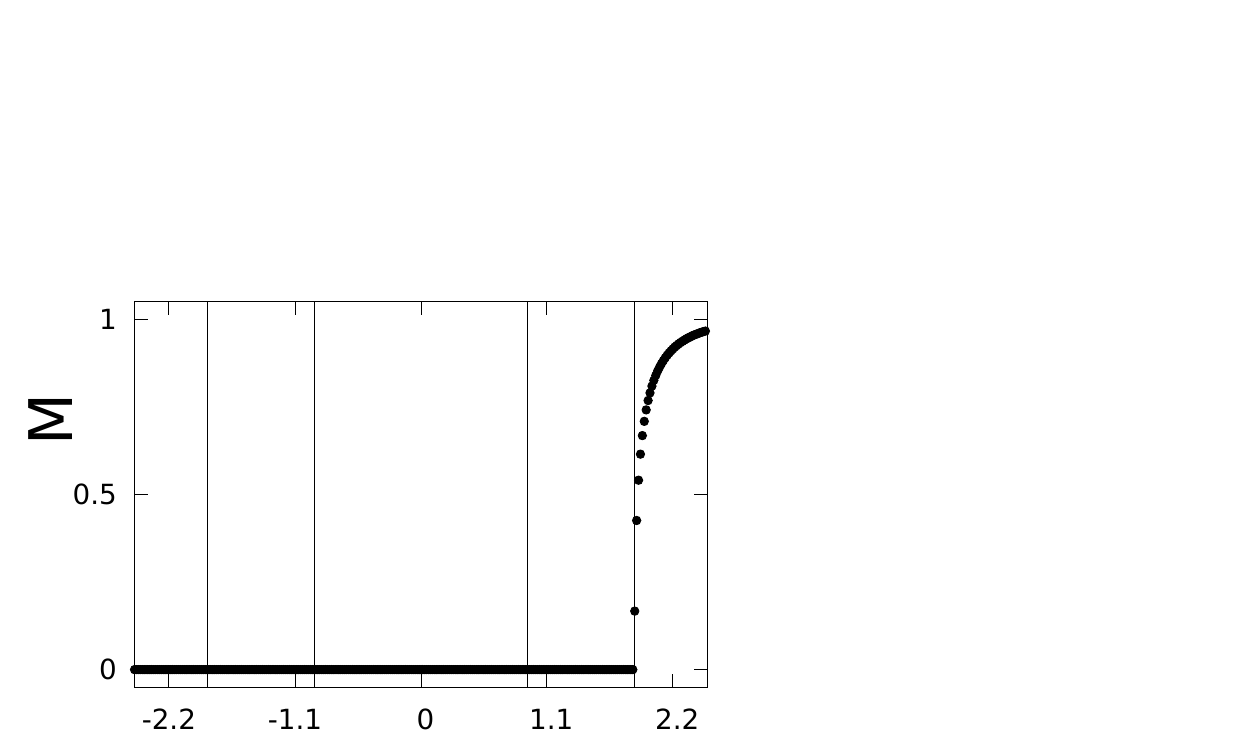}}}
}
\put(150,460)
{
\resizebox{10cm}{!}{\rotatebox{0}{\includegraphics{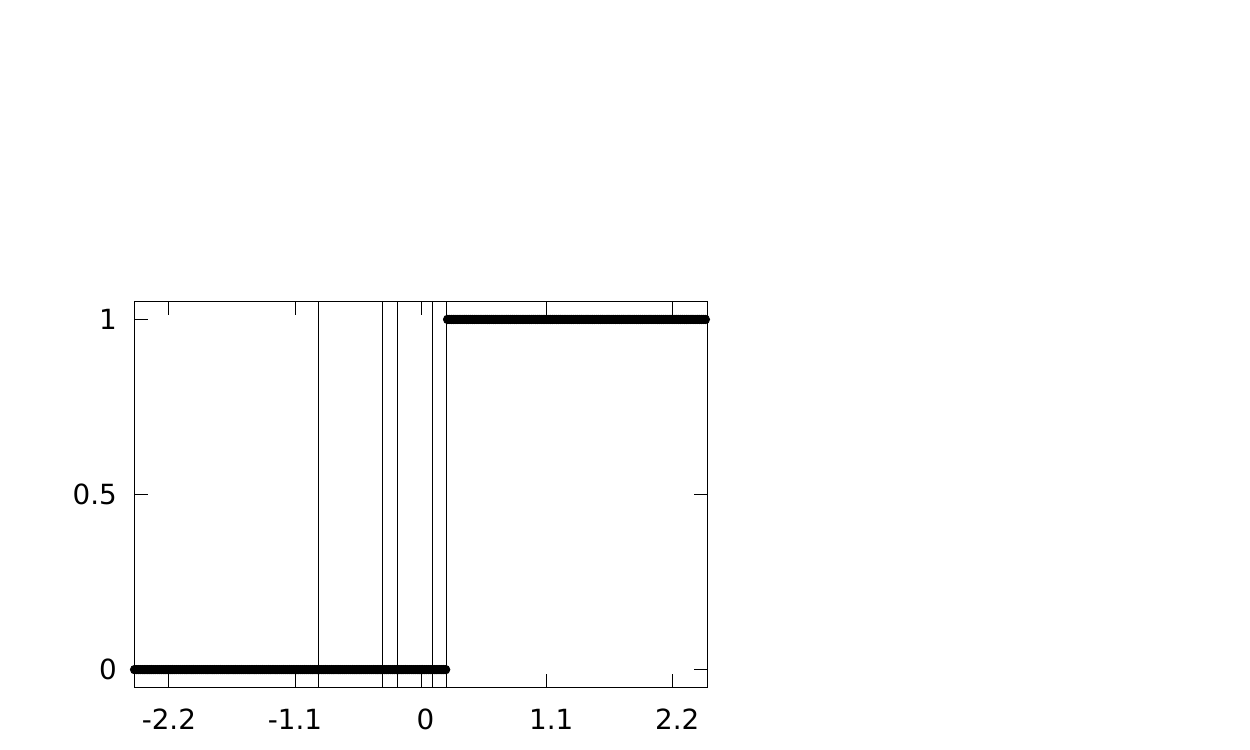}}}
}
\put(320,460)
{
\resizebox{10cm}{!}{\rotatebox{0}{\includegraphics{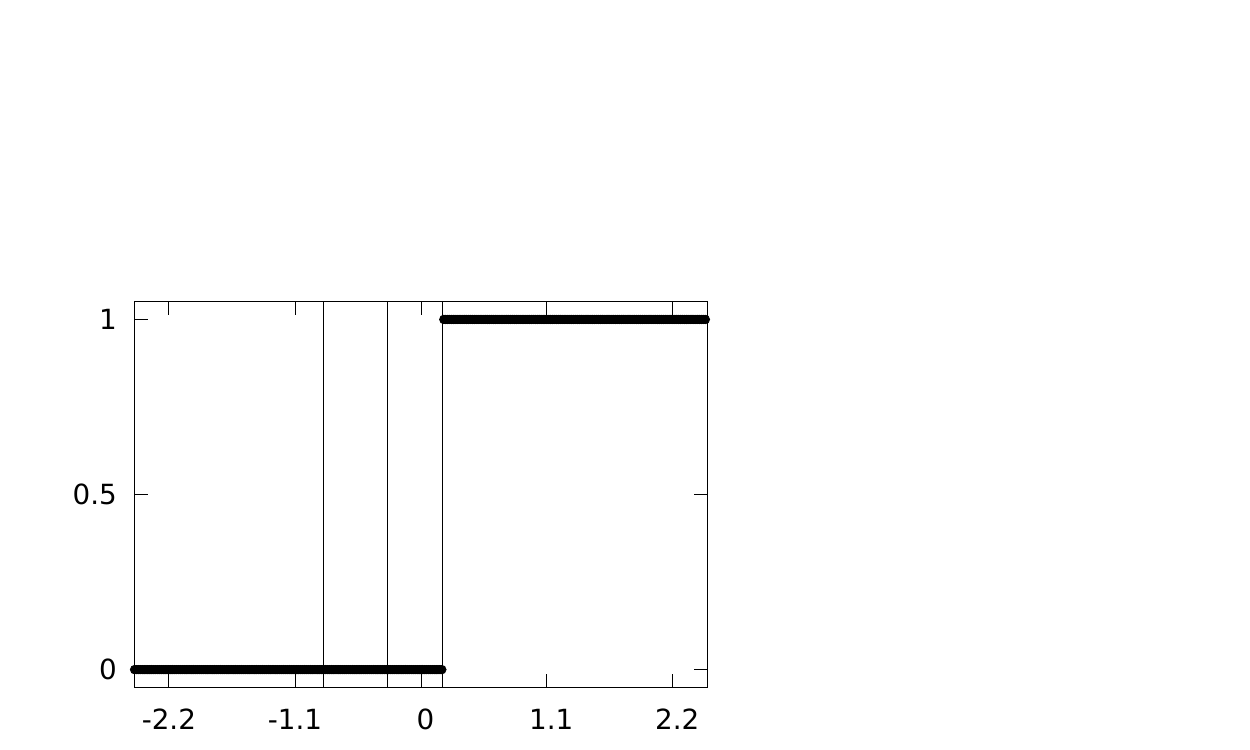}}}
}
\put(75,-10){$K$}
\put(245,-10){$K$}
\put(415,-10){$K$}
\end{picture}
\vskip 0.2 cm \caption{Order parameters vs.\ $K$ for different
values of the constraint parameter $\lambda$. From the left to the
right $\lambda$ takes the values $0$, $3.6$, and $6$. The thin
vertical lines indicate points where a phase transition occurs.
From the left to the right the phases are piled-up, tetrahedral,
disordered, flat--octahedral, and flat at $\lambda=0$, piled-up,
tetrahedral, disordered, piled--octahedral, disordered, and flat at
$\lambda=3.6$, piled-up, tetrahedral, piled--octahedral, and flat at
$\lambda=6$.
 }
\label{f:acdin05}
\end{figure*}
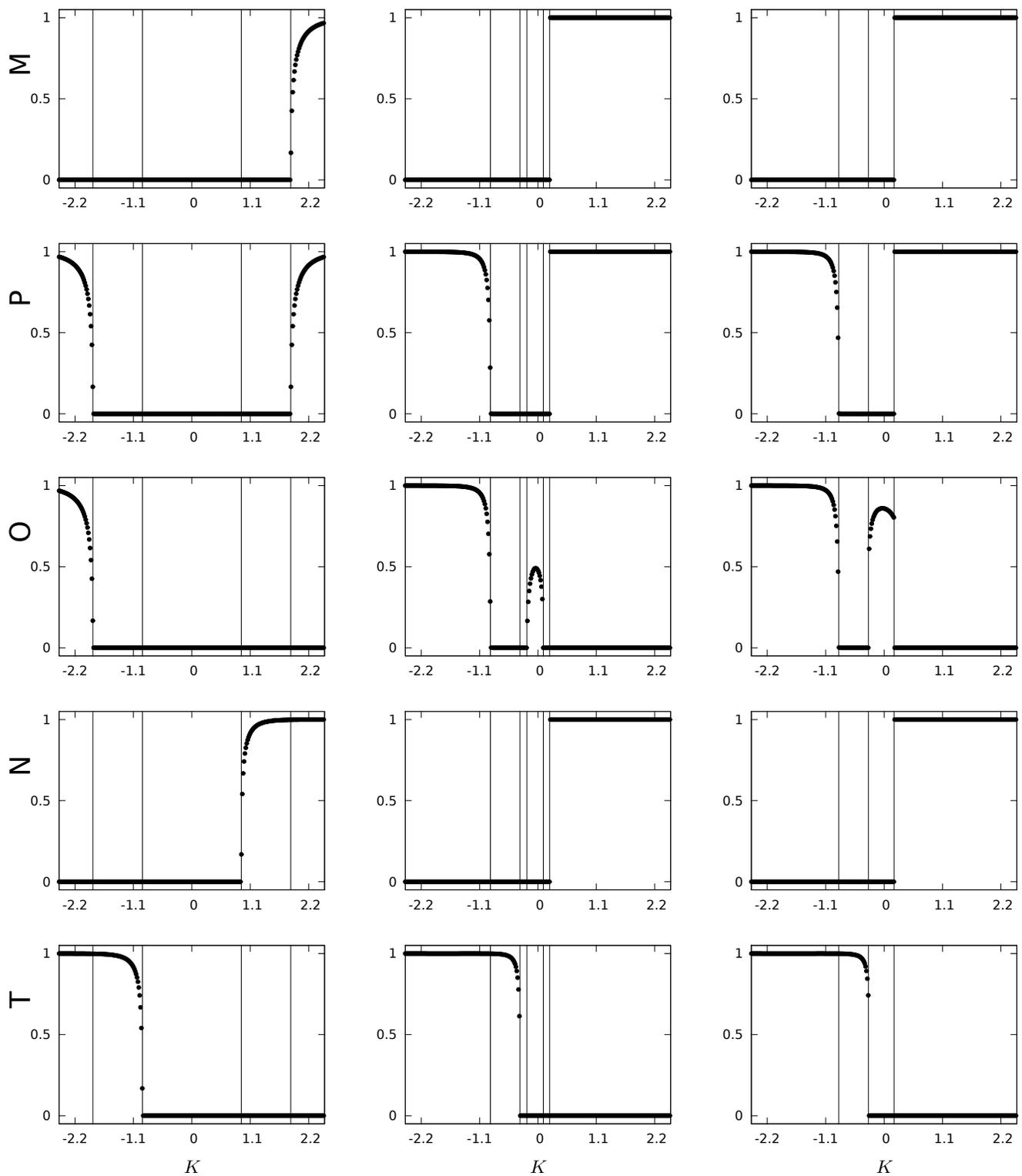

Moving to lower values of $\lambda$, for $\lambda \gtrsim 4.91$ the
behavior described above remains qualitatively the same as in the
case without defects, as it can be seen in Fig.~\ref{f:fase}. An
illustration, for $\lambda=6$, in terms of the order parameters as
functions of $K$, is reported in the right column of
Fig.~\ref{f:acdin05}.

By further decreasing the value of $\lambda$ a new phase,
characterized by the vanishing of all the order parameters, appears
in the phase diagram. This phase, which we shall call disordered, is
bounded by the tetrahedral and the p--octahedral phases at negative
$K$ and by the p--octahedral, the flat and the f--octahedral (see
below) phases at positive $K$. The p--octahedral--disordered and
tetrahedral--disordered phase transitions are continuous and the
corresponding critical lines meet the p--octahedral--tetrahedral
first--order transition line at a bicritical point at $K=-0.298$,
$\lambda=4.90$ (see Fig.~\ref{f:fase}). At positive $K$, the
flat--disordered transition is first--order, like the
flat--p--octahedral one, and the p--octahedral--disordered critical
line meets them at a critical end--point at $K=0.208$ and
$\lambda=3.84$.  At zero curvature, the p--octahedral--disordered
transition occurs at $\lambda=3.478$.  The behavior of the order
parameters as a function of $K$ at $\lambda=3.6$ is reported in the
central column of Fig.~\ref{f:acdin05}.

\begin{figure*}
\begin{picture}(500,300)
\put(-10,0) {
\includegraphics[width=10.5cm, height=8.5cm]{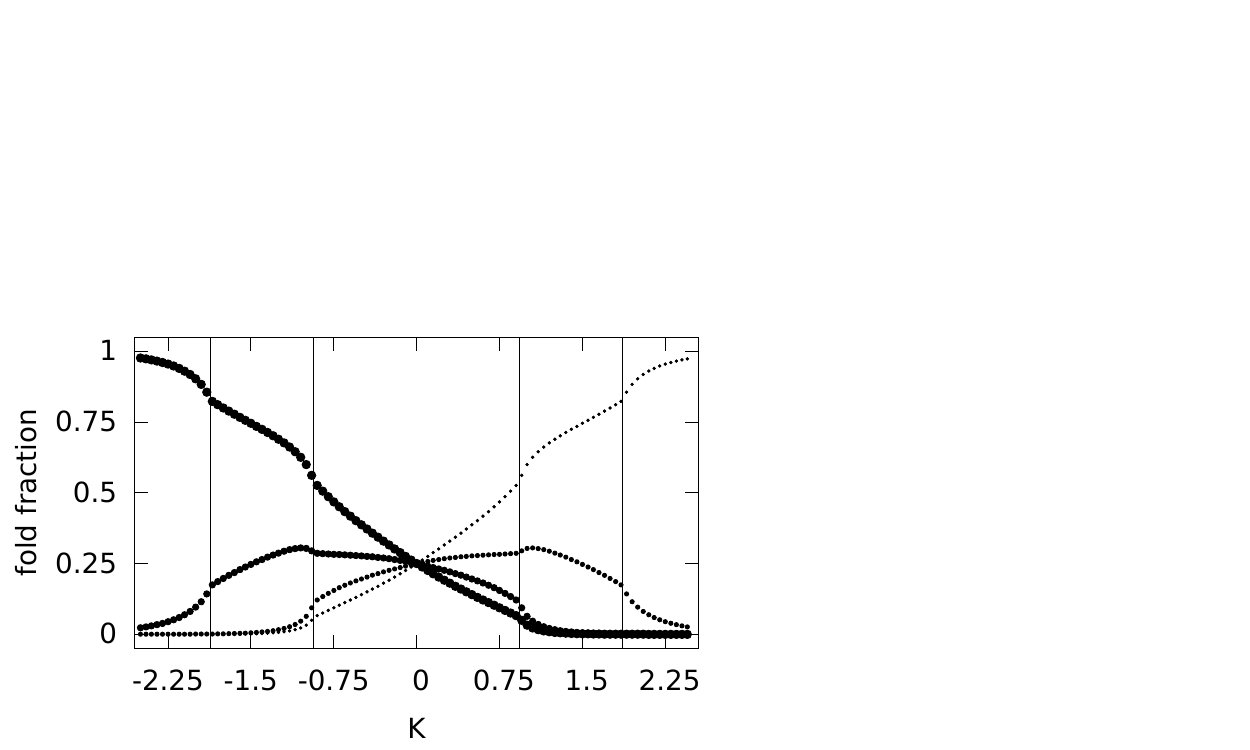}
}
\put(160,0) {
\includegraphics[width=10.5cm, height=8.5cm]{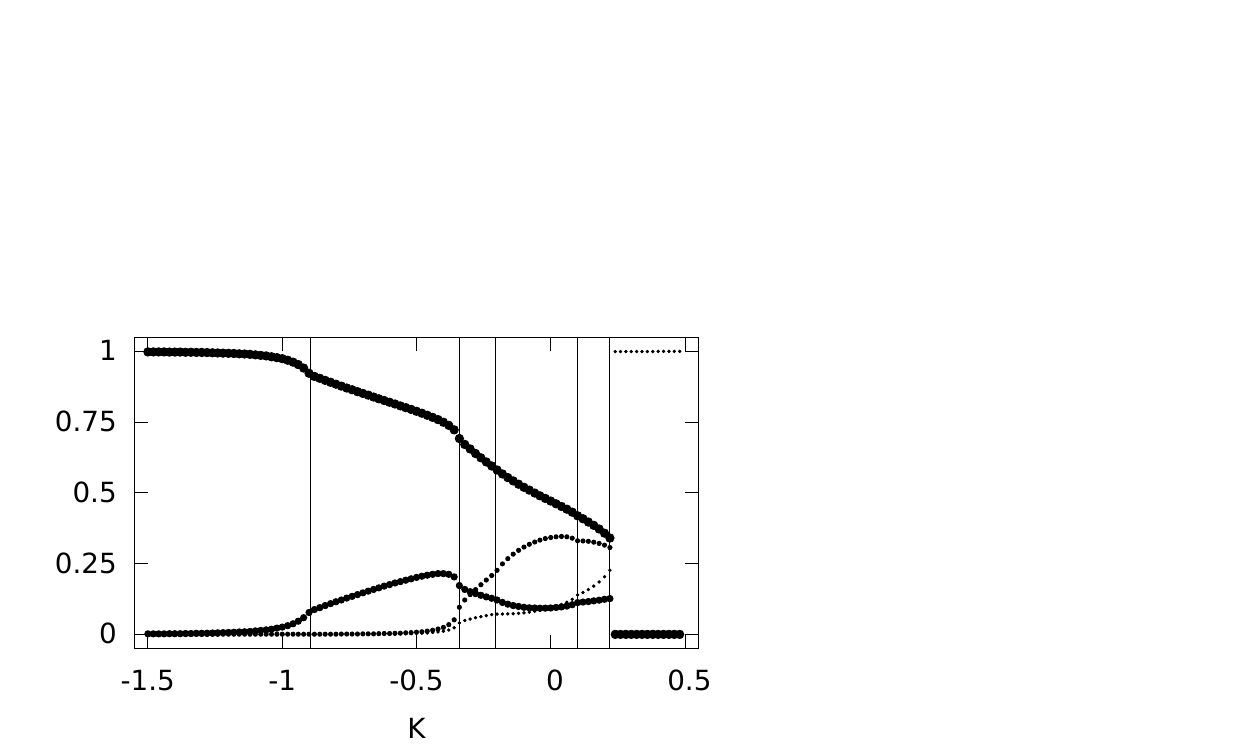}
}
\put(330,0) {
\includegraphics[width=10.5cm, height=8.5cm]{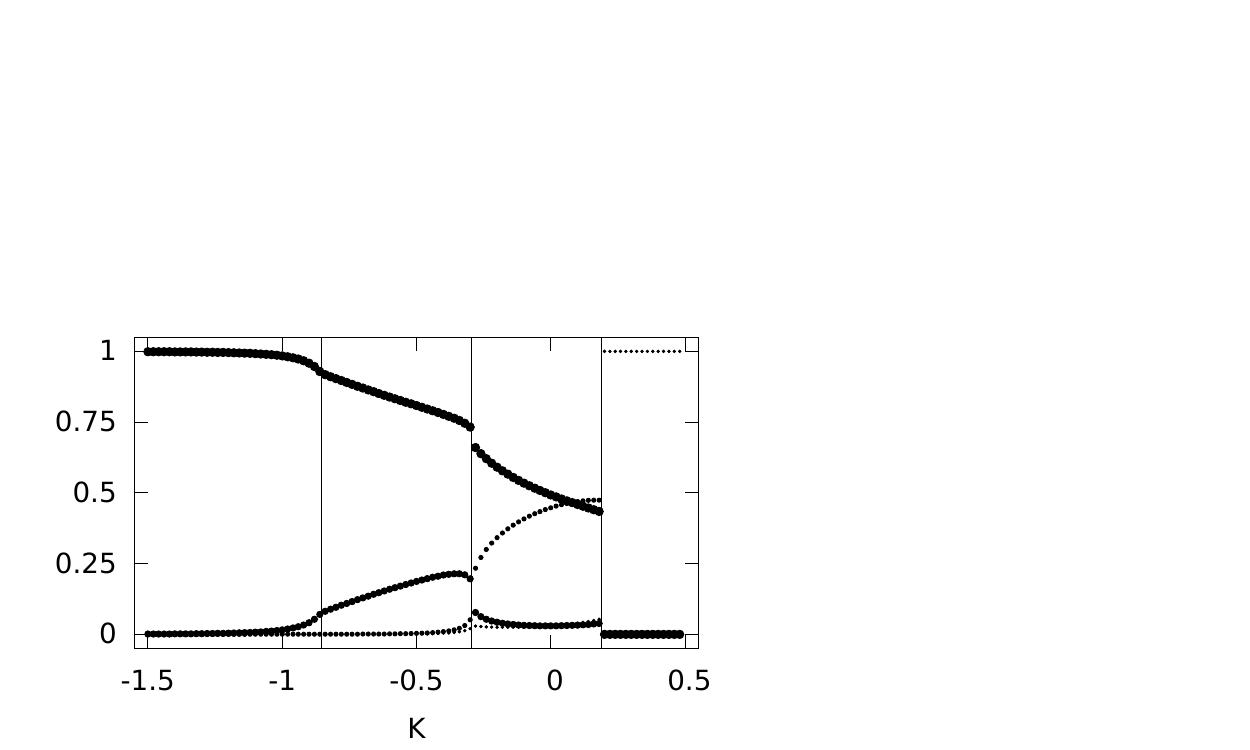}
}
\end{picture}
\vskip 0.2 cm \caption{Different fold  fractions. From the left to
the right $\lambda$ takes the values $0$, $3.6$, and $6$. The thin
vertical lines indicate points where a phase transition occurs.
From the left to the right the phases are piled-up, tetrahedral,
disordered, flat--octahedral, and flat at $\lambda=0$; piled-up,
tetrahedral, disordered, piled--octahedral, disordered, and flat at
$\lambda=3.6$; piled-up, tetrahedral, piled--octahedral, and flat at
$\lambda=6$. The dotted lines, in increasing thickness order,
denote respectively no fold, octahedral fold, tetrahedral fold,  and
complete fold. } \label{f:piegamenti}
\end{figure*}

Let us now proceed by describing our phase diagram at positive $K$
and small $\lambda$.  In this region another new phase, to be
denoted by f(lat)--octahedral, is found between the disordered and
the flat phase. The f--octahedral phase is characterized by
$\parametro{N}$ being the only non--vanishing order parameter, while
in terms of folds between adjacent plaquettes it exhibits a mixture
of no--folds and octahedral folds. This can be seen in
Fig.~\ref{f:piegamenti} where the average proportions of the four
types of folds is shown  as a function of $K$, for three values of
$\lambda$, $\lambda=0$, 3.6 and 6.

The  appearance of the f--octahedral phase can be better understood
by discussing the behavior of the model at $\lambda=0$. As observed
in \cite{BGGM}, on this axis the model reduces to the Ashkin--Teller
model \cite{Bax}, with parameters corresponding to the trivial case
of two independent Ising models. Starting from low $K$, two
continuous, symmetry breaking transitions are found. The first
transition, separating the disordered phase from the f--octahedral
phase, occurs at $K=3K_\rr{c}/2$, where $K_c$ is the critical Ising
coupling on the hexagonal lattice, as discussed in \cite{BGGM}. Here
the order parameter $\parametro{N}$ becomes different from zero. The
second transition, separating the f--octahedral phase from the flat
phase, occurs at $K=3K_\rr{c}$ \cite{BGGM}. Here the symmetries $z
\to -z$ and $\sigma \to -\sigma$ are separately broken, and the
order parameters $\parametro{M}$ and $\parametro{P}$ also become
different from zero.

We find, in our approximation, that the two transitions occur at
$K_{\rr{c}1} = 0.9321$ and at $K_{\rr{c}2}=1.8642$. These values
correspond to the CVM estimate of the Ising critical point
$K_\rr{c}=0.6214$ on the hexagonal lattice (in the hexagonal
plaquette approximation) \cite{nostro}. At positive $\lambda$ the
transitions separating the f--octahedral and the disordered phases
and the f--octahedral and the flat phases
 remain continuous. The
corresponding lines meet at a bicritical point at
$K=0.388$, $\lambda=2.39$ where the first--order disordered--flat
transition appears.

At negative $K$ the tetrahedral--piled--up phase transition remains
continuous for all values of $\lambda$. The transition line
intersects the horizontal axis at $K=-K_{\rr{c}2}$ where the
symmetries $z\to-z$ and $\sigma_\rr{st}\to-\sigma_\rr{st}$ are
separately broken. Moreover, at $\lambda=0$, the
disordered--tetrahedral continuous transition is found at
$K=-K_{\rr{c}1}$.  These critical values correspond to the
anti--ferromagnetic images of the Ising transitions occurring in the
equivalent Ashkin--Teller model at positive $K$.  The behavior of
the order parameters as a function of $K$ at $\lambda = 0$ is
reported in the left column of Fig.~\ref{f:acdin05}.

The above results show that the introduction of defects turns
first--order transition into continuous one, as in the planar
folding case \cite{nostro}. Here however the phase behavior is
richer and we can observe an additional effect due to the presence
of defects: the p--octahedral phase disappears, to be replaced by
the disordered phase and the f--octahedral phase, which also
exhibits a significant fraction of octahedral folds.

\begin{figure*}
\begin{picture}(300,400)
\put(-60,0)
{
\resizebox{24cm}{!}{\rotatebox{0}{\includegraphics{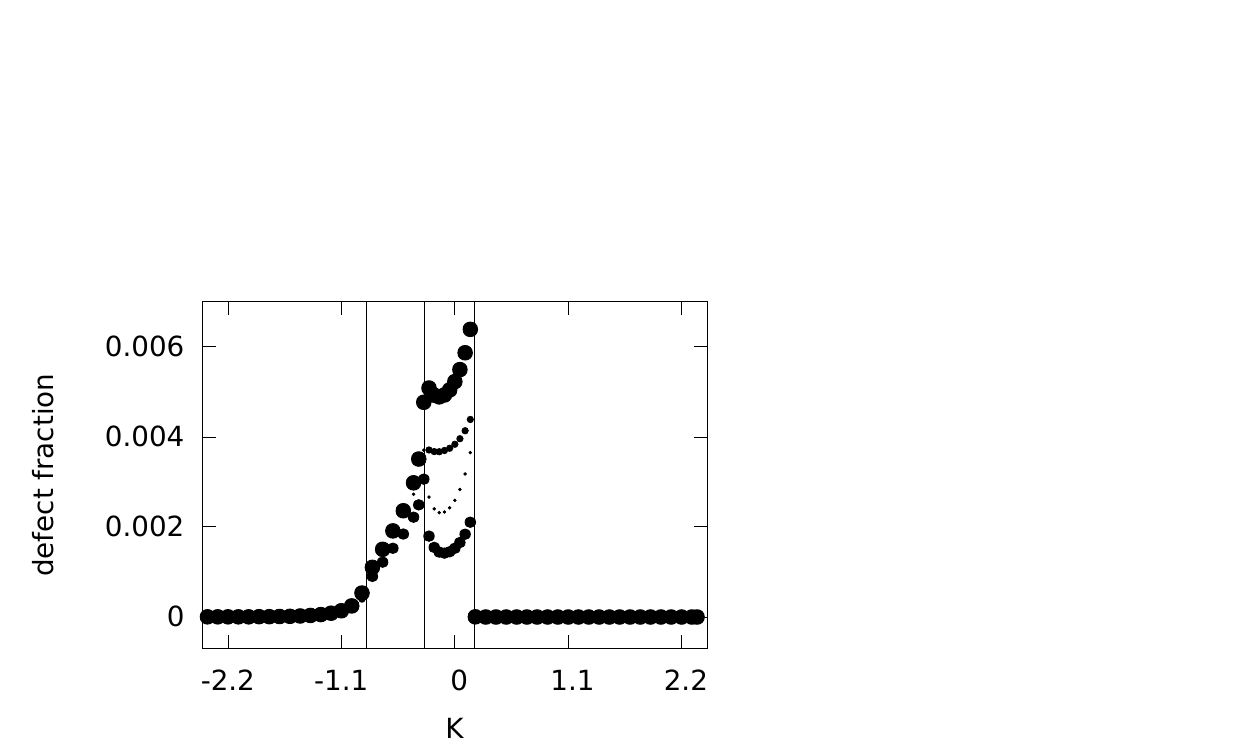}}}
}
\end{picture}
\caption{Fraction of defects as a function of $K$ for $\lambda=6$.
 The dotted lines, in increasing thickness order, denote respectively
 $p_L$, $p_{M_1}$, $p_{M_2}$, and $p$.
 The thin vertical lines indicate points where a phase transition occurs.
 From the left to the right the phases are
 piled-up, tetrahedral, piled--octahedral, and flat.
}
\label{f:difetti4}
\end{figure*}

\begin{figure*}
\begin{picture}(300,400)
\put(-60,0)
{
\resizebox{24cm}{!}{\rotatebox{0}{\includegraphics{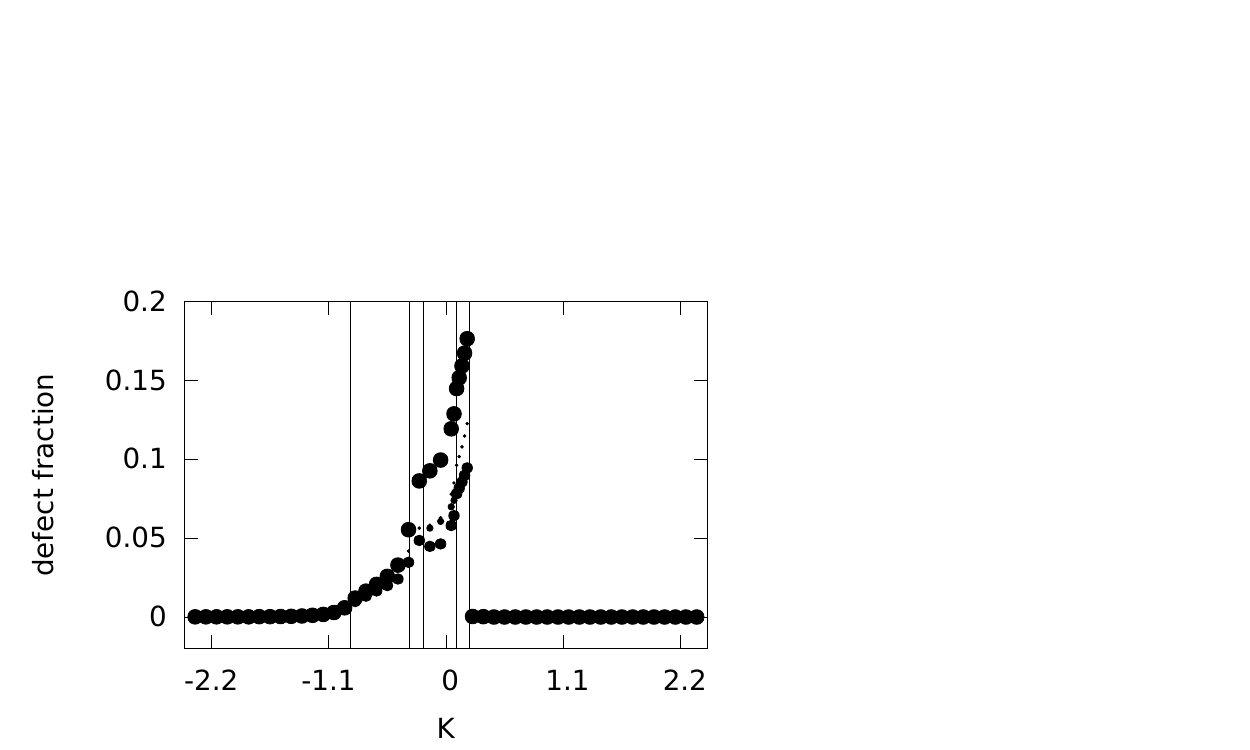}}}
}
\end{picture}
\caption{Same as Fig.\ \protect\ref{f:difetti4} for $\lambda=3.6$.
 From the left to the right the phases are
 piled-up, tetrahedral, disordered, piled--octahedral, disordered, and flat.
}
\label{f:difetti2}
\end{figure*}

\begin{figure*}
\begin{picture}(300,400)
\put(-60,0)
{
\resizebox{24cm}{!}{\rotatebox{0}{\includegraphics{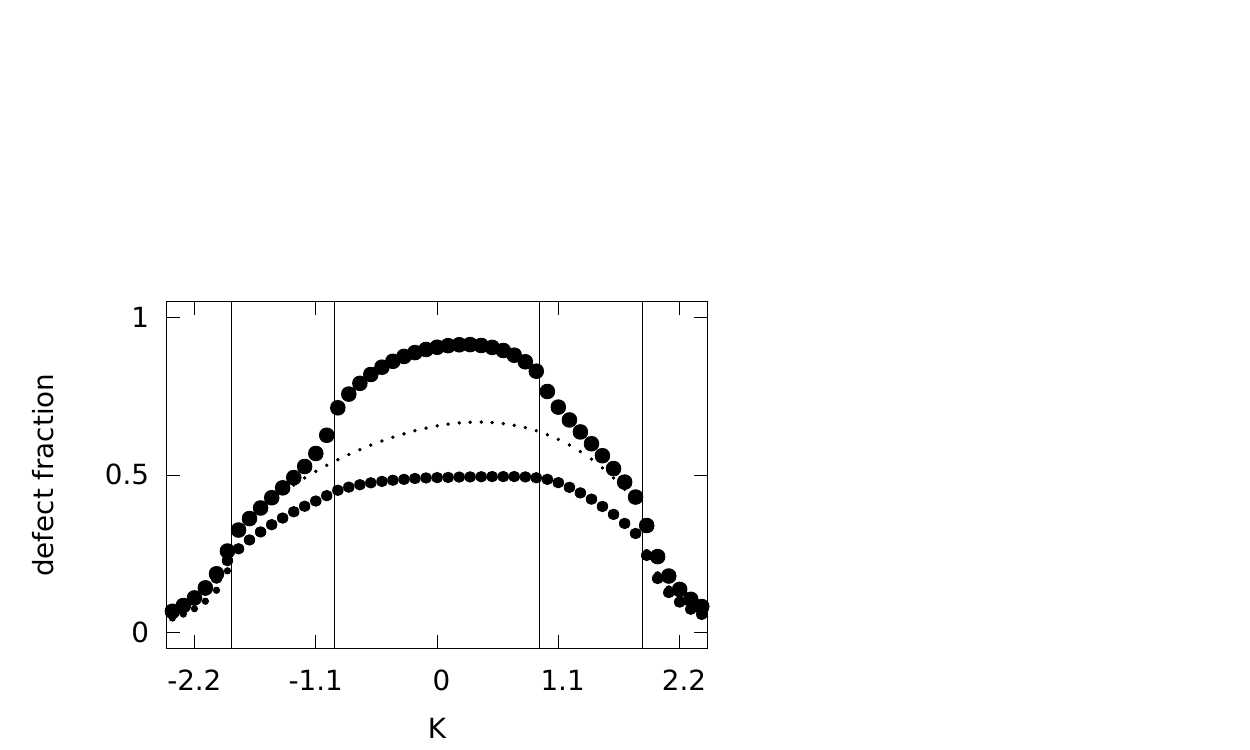}}}
}
\end{picture}
\caption{Same as Fig.\ \protect\ref{f:difetti4} for $\lambda=0$.
 From the left to the right the phases are
 piled-up, tetrahedral, disordered, flat--octahedral, and flat.
}
\label{f:difetti0}
\end{figure*}

It is also worth taking a look at the number of defects appearing in
our triangular lattice as a function of $K$ and $\lambda$. To be
more precise we define the fraction $p_L$ of hexagons at which
folding rule Eq.\ (\ref{rule1}) is violated
\begin{equation}
  p_L = \textrm{Tr}_6 \rho_6 \left( 1 - \fcar_{\{L(\sigma_e)=0\
      \textrm{mod}\ 3\}} \right),
\end{equation}
the fractions $p_{M_c}$ of hexagons at which folding rules
Eq.\ (\ref{rule2}) are violated
\begin{equation}
  p_{M_c} = \textrm{Tr}_6 \rho_6 \left( 1 -
    \fcar_{\{M_{c}(\sigma_e,z_e)=0\ \textrm{mod}\ 2\}} \right)
  \qquad c = 1, 2
\end{equation}
and the fraction $p$ of hexagons at which at least one folding rule is
violated,
\begin{equation}
\begin{array}{l}
 p = \textrm{Tr}_6 \rho_6 \Big(1
     -\fcar_{\{M_{c}(\sigma_e,z_e)=0\ \textrm{mod}\ 2\}}
     \times
\\
{\displaystyle
    \phantom{mmmmmmmmm}
    \times
    \prod_{c=1}^2 \fcar_{\{M_{c}(\sigma_e,z_e)=0\ \textrm{mod}\ 2\}}
\Big). }
\\
\end{array}
\end{equation}
These quantities are plotted as functions of $K$
in Fig.~\ref{f:difetti4} for $\lambda = 6$ (that is, close to the
defect--free limit),
in Fig.~\ref{f:difetti2} for $\lambda = 3.6$,
and in Fig.~\ref{f:difetti0} for $\lambda = 0$. We see that the
fraction of defects has a maximum close to $K = 0$ and decreases as
$|K|$ increases. Moreover, the fraction of defects is a decreasing
function of $\lambda$. Considering the various phases we see that
the phases exhibiting less fluctuations, namely the flat and
piled--up phases, are almost defect--free, even at small $\lambda$,
while the other phases are more prone to defects.

Notice also that $p_{M_1}$ and $p_{M_2}$ differ in the flat,
p--octahedral and piled--up phases. This difference between
$p_{M_1}$ and $p_{M_2}$ is a consequence of the breaking of the
global inversion symmetry $\sigma_i \to -\sigma_i, z_i \to -z_i,
\forall i$. This transformation maps configurations violating
(\ref{rule2}) for $c = 1$ to configurations violating (\ref{rule2})
for $c = 2$, and viceversa. The corresponding symmetry is preserved
only in the disordered, f--octahedral and tetrahedral phases, where
$p_{M_1} = p_{M_2}$ as a consequence.

\section{Conclusions}
\label{s:conclusioni}
\par\noindent

We have generalized a model for the folding transitions of a
triangular lattice in a three--dimensional space, discretized as a
fcc lattice, by allowing defects corresponding to cuts in the
triangular lattice, and weighing them by a suitable energy cost. We
have studied the model in a six--point approximation of the CVM. In
the limit of the energy cost of a defect going to $+\infty$, we
recover previous results for the defect--free model
\cite{nostro3d,BGGM}. On the other hand, when this energy cost is
sufficiently small, we find that first--order transitions are turned
into continuous one, and that the octahedral phase found in the
defect--free limit, which was characterized by octahedral and
complete folds between adjacent plaquettes, is replaced by a fully
disordered phase and another phase characterized by octahedral folds
and no--folds, which is related to the intermediate temperature
phase of the Ashkin--Teller model.  The model has a rich phase
diagram with several multicritical points, namely, two bicritical
points and a critical end--point.  We have also
shown that defects are more likely to occur in phases exhibiting
larger fluctuations, while the flat and piled--up phases are almost
defect free, and their concentration typically decreases as the
absolute value of the curvature energy increases.

$\phantom.$
\newpage

\newpage

\newpage

\newpage

\newpage

\newpage

\newpage




\begin{thebibliography}{00000}
\bibitem{Nota} We also studied a different  model where the defects
  corresponding to separate violations of the folding rules
  (\ref{rule1}) and (\ref{rule2}) (c=1,2) are weighted by distinct
  energy costs.
The  resulting  phase diagram is qualitatively very
  similar to that shown in Fig.\ \ref{f:fase}, the only qualitative
  change being the disappearance of a tricritical point on the flat--f--octahedral transition line.
\end{thebibliography}
\end{document}